\documentclass[preprint,aps,prd,nofootinbib,preprintnumbers,amsmath,amssymb]{revtex4}
\usepackage{graphicx}
\usepackage{dcolumn}
\usepackage{bm}
\usepackage{natbib}
\usepackage{color}
\usepackage{slashed}

\newcommand\fft[2]{\frac{#1}{#2}}
\newcommand\ft[2]{{\textstyle\frac{#1}{#2}}}
\newcommand{\btop}[2]{\genfrac{[}{]}{0pt}{}{\,#1\,}{\,#2\,}}
\newcommand{\nc}{\newcommand}
\newcommand\nn{{\nonumber}}
\newcommand{\beq}{\begin{equation}}
\newcommand{\eq}{\end{equation}}

\nc{\bea}{\begin{eqnarray}} \nc{\ea}{\end{eqnarray}} \nc{\be}{\begin{equation}} \nc{\ee}{\end{equation}} \nc{\barr}{\begin{array}}
\nc{\earr}{\end{array}}

\begin{document}

\preprint{MCTP-11-10}

\title{Supersymmetry of consistent massive truncations of IIB supergravity}

\author{James T.~Liu}
\email{jimliu@umich.edu}
\affiliation{Michigan Center for Theoretical Physics,
Randall Laboratory of Physics,
The University of Michigan,
Ann Arbor, MI 48109--1040, USA}

\author{Phillip Szepietowski}
\email{pszepiet@umich.edu}
\affiliation{Michigan Center for Theoretical Physics,
Randall Laboratory of Physics,
The University of Michigan,
Ann Arbor, MI 48109--1040, USA}

\begin{abstract}

We discuss the supersymmetry and fermionic sector of the recently obtained consistent truncations
of IIB supergravity containing massive modes. In particular, we present the
general form of the five-dimensional $\mathcal N = 4$ supersymmetry
transformations and equations of motion for the fermions arising in the reduction of IIB theory on
$T^{1,1}$ which contains all modes invariant under the $\rm SU(2)\times SU(2)$
isometry group.  The $\mathcal N=4$ reduction can be further truncated to
two different $\mathcal N = 2$ sub-sectors.  For each of these, we present
the $\mathcal N=2$ fermionic supersymmetry transformations and corresponding
superpotentials.  As an application, we obtain the explicit Killing spinors
of the Klebanov-Strassler solution and comment on the relation to the ansatz
of Papadopoulos and Tseytlin. We also demonstrate the applicability of
consistent truncations on squashed Sasaki-Einstein manifolds to a class of
flux compactifications, focusing on a recent solution describing the
geometry of gaugino condensation on wrapped D7 branes and which possesses
dynamic SU(2) structure.
\end{abstract}

\maketitle

\section{Introduction}

Kaluza-Klein reductions of maximal supergravity theories have received renewed attention in recent years. In particular, there has been much work on consistent reductions which incorporate only a finite number of massive supergravity modes in the dimensionally reduced theory. These constructions rely on decomposing the field content of either 11-dimensional or IIB supergravity in terms of singlets of the isometry or structure group of the internal manifold. The essential feature is that these singlet modes form a closed set when acted on by differential operators, thus guaranteeing the consistency of the reduction.

Initial breathing and squashing mode sphere reductions were carried out in
\cite{Bremer:1998zp,Liu:2000gk}, and subsequently extended in
\cite{Gauntlett:2009zw} to the case of squashed Sasaki-Einstein manifolds
in 11-dimensional supergravity yielding gauged $\mathcal N=2$ supergravity
coupled to complete supermultiplets in four-dimensions. The analogous case for IIB supergravity has since been worked out in \cite{Liu:2010sa,Gauntlett:2010vu,Cassani:2010uw,Skenderis:2010vz}, yielding five-dimensional $\mathcal N = 4$ gauged supergravity, with \cite{Liu:2010sa} emphasizing a sub-truncation to $\mathcal N=2$. These reductions retain only singlets of the SU(2) structure group on the Sasaki-Einstein manifold, which allows for a rather complicated decomposition yielding many massive fields while at the same time ensuring consistency of the reduction.

Recently, similar reductions of IIB supergravity on $T^{1,1}$ have been considered in \cite{Cassani:2010na,Bena:2010pr}\footnote{For a similar construction considering massive reductions of IIA supergravity on coset spaces see \cite{Cassani:2009ck}.}, which, due to the non-trivial cohomology and reduced structure group of $T^{1,1}$, allow for additional deformations. An important feature of these reductions is the generality of the ansatz, as they allow for all possible deformations which are invariant under the particular isometry or structure group. As noted in \cite{Bena:2010pr,Cassani:2010na} the reductions on $T^{1,1}$ actually include the entire content of the Papadopoulos-Tseytlin (PT) ansatz \cite{Papadopoulos:2000gj} as a sub-truncation, which itself includes many interesting solutions studied in the literature \cite{Klebanov:2000hb,Maldacena:2000yy,Pando Zayas:2000sq}. We will furthermore demonstrate the applicability of these reductions to recent flux compactifications which exhibit dynamic SU(2) structure \cite{Heidenreich:2010ad}.

The explicit knowledge of the singlet sector of the isometry group of the internal manifold also has benefits when considering the fermionic sector of the IIB theory, allowing one to work out explicitly both the fermionic terms in the Lagrangian and their supersymmetry transformations. The fermionic content and supersymmetry of the general reduction on Sasaki-Einstein manifolds has been analyzed in great detail for both 11-dimensional and IIB supergravity in \cite{Bah:2010yt,Bah:2010cu} and, again with an emphasis on the $\mathcal N = 2$ sector of the IIB reduction, in \cite{Liu:2010pq}.

In this paper we continue the study of the supersymmetry of these types of reductions for the case of IIB supergravity by explicitly working out the fermionic supersymmetry transformations and equations of motion for the general $T^{1,1}$ reductions in \cite{Bena:2010pr,Cassani:2010na}. We begin by presenting the fermionic reduction in the context of five-dimensional $\mathcal N = 4$ supergravity coupled to three vector multiplets. This includes reducing the supersymmetry transformations and the fermion equations of motion. We will then proceed to demonstrate appropriate truncations to five-dimensional $\mathcal N = 2$ matter coupled supergravity, in the process determining the $\mathcal N = 2$ superpotential from the supersymmetry equations. The superpotential in particular may be of interest for future studies and to the classification of supersymmetric solutions to these types of reductions.

Following a presentation of the details of the reduction, we comment on the relation to various supergravity solutions in the literature, focusing on the PT ansatz \cite{Papadopoulos:2000gj} and the Klebanov-Strassler (KS) solution \cite{Klebanov:2000hb}.  Demanding the vanishing of the fermionic supersymmetry transformations gives rise to a set of Killing spinor equations for these backgrounds.  Using this five-dimensional supergravity approach, we explicitly
construct the Killing spinors of the KS solution and also comment on the
relevant superpotential.

Finally, although we focus on five-dimensional gauged supergravity, the
supersymmetry analysis is also applicable to a wide range of solutions
involving Calabi-Yau cones.  As an example, we
discuss the recent construction of \cite{Heidenreich:2010ad} which describes a solution of IIB theory corresponding to wrapped D7-branes on a $\mathbb{CP}^2$ cone with nontrivial imaginary anti-self-dual (IASD) flux. As discussed in \cite{Koerber:2007xk,Heidenreich:2010ad,Baumann:2010sx}, such IASD flux is sourced by a gaugino bilinear on the world volume of the D7-branes, and so this background is considered to be a supergravity description of gaugino condensation.\footnote{For recent related developments on this also see \cite{Dymarsky:2010mf}.} In addition, the $\mathbb{CP}^2$ cone possesses a dynamic SU(2) structure. Here, we will describe the relation of the bosonic solution to the massive truncations on squashed Sasaki-Einstein manifolds and outline the supersymmetry conditions in this language. It is important to note that \cite{Heidenreich:2010ad} includes a detailed discussion of the supersymmetry conditions. In particular, \cite{Heidenreich:2010ad} relies on the pure spinor formalism to express the susy conditions and our presentation will be complementary to the one presented in \cite{Heidenreich:2010ad}. We also comment on the manifestation of the dynamic SU(2) structure within the present reduction framework.

In the next section, we begin with a review of the reduction of IIB
supergravity on squashed Sasaki-Einstein manifolds and on the deformed
$T^{1,1}$. In Section \ref{sec:N4}, we reduce the IIB gravitino and dilatino
transformations and equations of motion on $T^{1,1}$ to obtain the corresponding equations of
the resulting five-dimensional $\mathcal N=4$ fermions. This is followed in section \ref{sec:N2}
by a further truncation of the supersymmetry transformations into two possible
$\mathcal N=2$ sub-sectors.  We
apply the resulting Killing spinor equations in Sections~IV and V to the
PT ansatz and wrapped D7-brane solution of \cite{Heidenreich:2010ad},
respectively.  We then conclude in Section~VI by a brief discussion of
future avenues to explore.

The reader interested only in the results can refer directly to equations (\ref{eq:susydil})--(\ref{eq:N4gravsusy}) for the complete $\mathcal N = 4$ supersymmetry variations and to equations (\ref{eq:N4dileom}), (\ref{eq:psi5eom})--(\ref{eq:psi9eom}) and (\ref{eq:N4graveom}) for the full $\mathcal N = 4$ equations of motion. The corresponding bosonic lagrangian%
\footnote{Note that, to match notations and conventions, one must also compare the bosonic reduction ansatz we present in Section \ref{sec:red} with those presented in \cite{Cassani:2010na} and \cite{Bena:2010pr}. The comparison with the latter is given in \cite{Halmagyi:2011yd}.}
can be found in Section 3.3 of \cite{Cassani:2010na} and Section 2.2 of \cite{Bena:2010pr}. Additionally, if interested only in the $\mathcal N = 2$ truncations, the supersymmetry variations following from the bosonic truncations discussed in Section 7.1  of \cite{Cassani:2010na} can be found in equations (\ref{eq:BVsusy1}), (\ref{eq:BVsusy2}) and (\ref{eq:BHsusy}).

\section{IIB supergravity on squashed $T^{1,1}$}\label{sec:red}

The reduction of the bosonic sector of IIB supergravity on a squashed Sasaki-Einstein manifold $SE_5$ has been carried out in \cite{Cassani:2010uw,Liu:2010sa,Gauntlett:2010vu,Skenderis:2010vz}, utilizing the SU(2) structure group. Similar technology was further employed to reduce IIB theory on a squashed and twisted $T^{1,1}$ \cite{Cassani:2010na,Bena:2010pr}, where the reduced U(1) structure group allowed for the introduction of one extra hyper-multiplet and vector
multiplet to the universal field content on $SE_5$. Here we begin by summarizing the bosonic reduction on $T^{1,1}$ presented in \cite{Cassani:2010na,Bena:2010pr}. We follow the IIB supergravity and reduction conventions of \cite{Liu:2010pq}.

The reduction follows by decomposing the IIB field content on U(1) invariant forms along $T^{1,1}$. We begin by writing the metric on $T^{1,1}$ in terms of the complex vielbein, $E_1$ and $E_2$, on the $\mathbb{CP}^1\times\mathbb{CP}^1$ base and a fibre direction $g_5$,
\begin{equation}
ds_{10}^2=e^{2A}ds_5^2+\ft16e^{2B_1}E_1 \bar{E}_1
+\ft16e^{2B_2}\hat E_2\hat{\bar E}_2+\ft19e^{2C}(g_5+3A_1)^2.
\end{equation}
Here we have defined the shifted vielbein,
\begin{equation}
\hat E_2=E_2+\alpha \bar E_1,
\end{equation}
where $\alpha$ is a complex function of the spacetime coordinates.  We furthermore fix $3A + 2B_1 + 2B_2 + C = 0$ to remain in the Einstein frame; this constraint should be assumed in all subsequent expressions.

For the form fields, we decompose the internal components along the invariant forms,
\begin{equation}
J_+ \equiv J_1+J_2, \qquad J_-\equiv J_1-J_2, \qquad \Omega, \qquad \eta = \ft13 g_5,
\end{equation}
where $J_1$ and $J_2$ denote the (untwisted) Kahler forms on each $\mathbb{CP}^1$ factor on the base and $J_+$ is the Kahler form on the entire base $\mathbb{CP}^1\times\mathbb{CP}^1$.  Explicitly, these are given in terms of the vielbein by
\begin{equation}
J_1=\fft{i}{12}E_1\wedge\bar E_1,\qquad
J_2=\fft{i}{12}E_2\wedge\bar E_2,\qquad
\Omega=\fft16E_1\wedge E_2.
\end{equation}
These forms also satisfy the structure relations
\begin{eqnarray}
J_1 \wedge \Omega = 0, & J_2\wedge\Omega = 0 & \Omega\wedge\bar\Omega = 4 J_1\wedge J_2 = 4 *_4 1, \nonumber \\
*_4 J_1 = J_2, &\qquad *_4 J_2 = J_1, \qquad &*_4\Omega = \Omega,
\end{eqnarray}
along with the differential relations
\begin{equation}
d\eta = 2J_+, \qquad dJ_1 = 0,\qquad dJ_2=0, \qquad d\Omega = 3i\Omega\wedge\eta.
\end{equation}
For generic $SE_5$, the two-form $J_-$ is no longer an invariant of the structure group and $J \equiv J_+$ along with $\Omega$ encode the SU(2) structure of the Sasaki-Einstein manifold.

For the three-form, we expand the two form potentials as
\begin{equation}
B_2^i=b_2^i+b_1^i\wedge(\eta+A_1)+c_0^iJ_++e_0^iJ_-+2\Re(b_0^i\Omega),
\end{equation}
and write
\begin{equation}
F_3^i=dB_2^i+j_0^iJ_-\wedge(\eta+A_1),
\end{equation}
where $j_0^i$ is a constant flux term.  Explicitly, for the three forms, we have
\begin{equation}
F_3^i=g_3^i+g_2^i\wedge(\eta+A_1)+(g_1^i+h_1^i)\wedge J_1
+(g_1^i-h_1^i)\wedge J_2+j_0^iJ_-\wedge(\eta+A_1)+2\Re[f_1^i\wedge\Omega
+f_0^i\Omega\wedge(\eta+A_1)],
\end{equation}
where
\begin{eqnarray}
&g_3^i=db_2^i-b_1^i\wedge F_2,\qquad g_2^i=db_1^i,\qquad
g_1^i=dc_0^i-2b_1^i,&\nonumber\\
&h_1^i=de_0^i-j_0^iA_1,\qquad f_1^i=Db_0^i,\qquad f_0^i=3ib_0^i.&
\end{eqnarray}
Here, $D=d-qiA_1$ is the U(1) gauge covariant derivative, and $q=3$ for
$b_0^i$. For the five-form, we expand the field strength explicitly
\begin{eqnarray}
\widetilde F_5&=&(1+*)[(4+\phi_0)*_41\wedge(\eta+A_1)+\mathbb A_1\wedge*_41
+p_{21}\wedge J_1\wedge(\eta+A_1)\nonumber\\
&&\kern3em
+p_{22}\wedge J_2\wedge(\eta+A_1)+2\Re(q_2\wedge\Omega\wedge(\eta+A_1))],
\end{eqnarray}
where we can split $p_{21}$ and $p_{22}$ into components $p_2$ and $r_2$ along $J_+$ and $J_-$, respectively, as
\begin{equation}
p_{21} = p_2 + r_2, \qquad p_{22} = p_2 - r_2.
\end{equation}

For the reduction of the supersymmetry transformations it is useful to define shifted forms, defined in terms of the shifted vielbein,
\begin{equation}
\hat J_1=\fft{i}{12}\hat E_1\wedge\hat{\bar E}_1,\qquad
\hat J_2=\fft{i}{12}\hat E_2\wedge\hat{\bar E}_2,\qquad
\hat\Omega=\fft16\hat E_1\wedge\hat E_2.
\end{equation}
The detailed form of the shifted variables is given in
Appendix~\ref{sec:shiftedapp}.
Finally, the five-form equation of motion imposes the following constraints
\begin{equation}
\phi_0=-6i\epsilon_{ij}(b_0^i\bar b_0^j-\bar b_0^ib_0^j)+\epsilon_{ij}
(j_0^ie_0^j-j_0^je_0^i),
\end{equation}
along with
\begin{equation}
d\mathbb A_1+2p_{21} +2p_{22} + 4F_2=d[2\epsilon_{ij}(b_0^iD\bar b_0^j + \bar b_0^iDb_0^j)-\epsilon_{ij}
e_0^i(h_1^j-j_0^jA_1)] + \epsilon_{ij}g_1^i\wedge g_1^j.
\end{equation}
Note that, as previously mentioned, the above expansions will yield a consistent reduction on $T^{1,1}$ \cite{Cassani:2010na,Bena:2010pr}. The reduction on
a generic Sasaki-Einstein manifold
\cite{Liu:2010sa,Gauntlett:2010vu,Cassani:2010uw,Skenderis:2010vz}
follows by setting
\begin{equation}
\{e^i_0, j^i_0, \alpha, r_2\} = 0, \qquad B_2 = B_1.
\end{equation}
This corresponds to removing the $\mathcal N=4$ Betti vector multiplet available
on $T^{1,1}$ but not in the generic case.

\section{The $\mathcal N = 4$ theory} \label{sec:N4}

As discussed in \cite{Cassani:2010na,Bena:2010pr}, the full reduction on $T^{1,1}$ is described by $\mathcal N = 4$ gauged supergravity in five-dimensions coupled to three vector multiplets. After a discussion of the reduction ansatz for the fermions, we present their supersymmetry transformations and equations of motion for the full $\mathcal N = 4$ theory.

\subsection{IIB spinor decomposition}

We take the IIB fermions to be complex Weyl spinors. The ten-dimensional
dilatino and gravitino variations are then given by \cite{Schwarz:1983qr}
\begin{eqnarray}\label{eq:susyvars}
\delta\lambda&=&\fft{i}{2\tau_2}\Gamma^M\partial_M\tau\epsilon^c
-\fft{i}{24}\Gamma^{MNP}v_iF^i_{MNP}\epsilon,\nonumber\\
\delta\Psi_M&=& \mathcal D_M \epsilon \nn \\
&=& (\nabla_M+\fft{i}{4\tau_2}\partial_M\tau_1+\fft{i}{16\cdot5!}
\Gamma^{NPQRS}\tilde F_{NPQRS}\Gamma_M)\epsilon
+\fft{i}{96}(\Gamma_M{}^{NPQ}-9\delta_M^N\Gamma^{PQ})v_iF^i_{NPQ}\epsilon^c.
\nonumber\\
\end{eqnarray}
Additionally, the type IIB fermion equations of motion are
\begin{eqnarray}
0&=&\Gamma^M\mathcal D_M\lambda-\fft{i}{8\cdot5!}\Gamma^{MNPQR}F_{MNPQR}
\lambda, \nn \\
0&=&\Gamma^{MNP}\mathcal D_N\Psi_P
+ \fft{i}{48}\Gamma^{NPQ}\Gamma^M
v_i^*F^{i*}_{NPQ}\lambda - \fft{i}{4\tau_2}\Gamma^N\Gamma^M\partial_N
\tau\lambda^c,
\label{eq:iibeom}
\end{eqnarray}
where the supercovariant derivative acting on the gravitino is defined in
the gravitino variation (\ref{eq:susyvars}).  The
supercovariant derivative acting on the dilatino takes the form
\begin{equation}
\mathcal D_M\lambda=\left(\nabla_M + \fft{3i}{4\tau_2}\partial_M\tau_1\right)
\lambda
- \fft{i}{2\tau_2}\Gamma^N\partial_N\tau\Psi_M^c
+ \fft{i}{24}\Gamma^{NPQ}v_iF^i_{NPQ}\Psi_M.
\end{equation}

Following \cite{Bah:2010yt,Bah:2010cu,Liu:2010pq}, we decompose the
ten-dimensional supersymmetry transformation parameter
$\epsilon$ in terms of the killing spinor of the internal Sasaki-Einstein manifold $\eta$ and its charge conjugate $\eta^c$ according to
\begin{equation}
\epsilon=e^{A/2}\left(\varepsilon_1\otimes\eta+\varepsilon_2\otimes\eta^c\right)
\otimes\btop10.
\end{equation}
As discussed in appendix \ref{app:covderiv}, the killing spinor $\eta$ satisfies the projections
\begin{eqnarray}
&(J_1)_{ab}\tilde\gamma^{b}\eta = i\tilde\gamma_a\eta,\qquad
(J_2)_{ab}\tilde\gamma^{b}\eta = i\tilde\gamma_a\eta,\qquad
\tilde\gamma^9\eta = -\eta, &\nonumber \\
&\Omega_{ab}\tilde\gamma^{b}\eta = -2\tilde\gamma_a\eta^c,\qquad
\bar\Omega_{ab}\tilde\gamma^{b}\eta = 0,&
\label{eq:etaproj}
\end{eqnarray}
and $\eta^c$ satisfies the charge conjugates of the above relations.
The conjugate spinor in ten-dimensions is then%
\footnote{We are using the gamma matrix conventions in \cite{Liu:2010pq}
with the exception that there is a relative minus sign in the
five-dimensional spacetime charge conjugation between the present work and
\cite{Liu:2010pq}.}
\begin{equation}
\epsilon^c\equiv - \Gamma_0 C_{10} \epsilon^* = ie^{A/2}\left(\varepsilon_2^c\otimes\eta-\varepsilon_1^c\otimes\eta^c\right)
\otimes\btop10.
\end{equation}

The two five-dimensional Dirac spinors $\varepsilon_i$ correspond to
four symplectic-Majorana spinors transforming under the $\mathbf 4$ of
Sp(4).  However, for simplicity, we retain a Dirac notation and treat
$\varepsilon_i$ as a doublet under an appropriate Sp(2) subgroup of the
full Sp(4) $R$-symmetry group.  In this case, we write the five-dimensional
spinors explicitly as a doublet under Sp(2) according to
\begin{equation}\label{eq:susyspinor}
\varepsilon = \btop{\varepsilon_1}{\varepsilon_2},
\end{equation}
where the upper component denotes the factor multiplying the killing spinor
$\eta$, while the lower component is the factor multiplying $\eta^c$.  Here, we
have suppressed the ten-dimensional chirality eigenvector $\btop10$.
In this notation, the action of the internal Dirac matrices on $\eta$ and
$\eta^c$ given in (\ref{eq:etaproj}) maps onto the five-dimensional spinors
$\varepsilon$ according to
\begin{eqnarray}
&\tilde\gamma^{b}(J_1)_{ab} \varepsilon = i\tilde\gamma_a\sigma_3\varepsilon, \qquad \tilde\gamma^{b}(J_2)_{ab} \varepsilon = i\tilde\gamma_a\sigma_3\varepsilon, \qquad \tilde\gamma^9\varepsilon = -\varepsilon &\nonumber \\
&\tilde\gamma^{b}\Omega_{ab}\varepsilon = -2\tilde\gamma_a\sigma_+\varepsilon,\qquad \tilde\gamma^{b}\bar\Omega_{ab}\varepsilon = 2\tilde\gamma_a\sigma_-\varepsilon.&
\end{eqnarray}
The $\sigma_i$ (with $\sigma_{\pm} = (\sigma_1 \pm i \sigma_2)/2$)
are standard Pauli matrices which act on the Sp(2) components of $\varepsilon$.

We expand the IIB fermions in a similar fashion. For the gravitino, we have
\begin{eqnarray}
&&\Psi_\alpha = e^{-A/2}\left(\psi_{1\,\alpha} \otimes\eta + \psi_{2\,\alpha}\otimes\eta^c\right)\otimes \btop{1}{0}, \nonumber \\
&&\Psi_a = e^{-A/2}\left(\psi^{(5)}_1 \otimes\tilde\gamma_a\eta + \psi^{(5)}_2\otimes\tilde\gamma_a\eta^c\right)\otimes \btop{1}{0}, \nonumber \\
&&\Psi_{\tilde a} = e^{-A/2}\left(\psi^{(7)}_1 \otimes\tilde\gamma_{\tilde{a}} \eta + \psi^{(7)}_2\otimes\tilde\gamma_{\tilde{a}}\eta^c\right)\otimes \btop{1}{0}, \nonumber \\
&&\Psi_9 = e^{-A/2}\left(\psi^{(9)}_1 \otimes\tilde\gamma_9\eta + \psi^{(9)}_2\otimes\tilde\gamma_9\eta^c\right)\otimes \btop{1}{0}, \nonumber \\
\end{eqnarray}
where $a = \{5,6\}$ and $\tilde a = \{7,8\}$. For the dilatino, we have
\begin{equation}
\lambda = e^{-A/2}\left(\lambda_1 \otimes\eta + \lambda_2\otimes\eta^c\right)\otimes \btop{0}{1}.
\end{equation}
These can also be expressed in an Sp(2) notation such that,
\begin{eqnarray}
&& \Psi_\alpha = e^{-A/2}\btop{\psi_{1\,\alpha}}{\psi_{2\,\alpha}}, \qquad
\Psi_a = e^{-A/2}\tilde\gamma_a\btop{\psi^{(5)}_1}{\psi^{(5)}_2}, \qquad
\Psi_{\tilde a} = e^{-A/2}\tilde\gamma_{\tilde a}\btop{\psi^{(7)}_1}{\psi^{(7)}_2}, \nonumber \\
&& \kern6em \Psi_9 = e^{-A/2}\tilde\gamma^9\btop{\psi^{(9)}_1}{\psi^{(9)}_2}, \qquad
\lambda = e^{-A/2}\btop{\lambda_1}{\lambda_2}.
\end{eqnarray}

\subsection{$D=5$, $\mathcal N = 4$ supersymmetry transformations}
\label{sec:N4susy}

As mentioned previously, the general reduction on $T^{1,1}$ yields $\mathcal N = 4$ supergravity in five-dimensions coupled to three vector multiplets \cite{Cassani:2010na,Bena:2010pr}. In later sections we will focus on further $\mathcal N = 2$ truncations, but here we first present the full $\mathcal N = 4$ fermion supersymmetry transformations.

Beginning with the dilatino, we arrive at
\begin{eqnarray}\label{eq:susydil}
\delta\lambda&=&-\fft{i}{2\tau_2}\gamma\cdot\partial\tau
\sigma_2\varepsilon^c-iv_i\biggl[\fft1{24}(e^{-2A}\gamma\cdot\hat g_3^i
-3ie^{-A-C}\gamma\cdot\hat g_2^i)\nonumber\\
&&+\fft{i}4\left(\gamma\cdot(e^{-2B_1}\hat g_{11}^i+e^{-2B_2}\hat g_{12}^i)
-ie^{A-C}(e^{-2B_1}\hat j_{01}^i+e^{-2B_2}\hat j_{02}^i)\right)
\sigma_3\nonumber\\
&&-e^{-B_1-B_2}\left(\gamma\cdot(\hat f_1^i\sigma_+
-\hat{\bar f}_1^i\sigma_-)
-ie^{A-C}(\hat f_0^i\sigma_+-\hat{\bar f}_0^i\sigma_-)\right)\biggr]\varepsilon.
\end{eqnarray}
The variation of the internal components of the gravitino gives
\begin{eqnarray}
\delta\psi^{(5)}&=&\biggl[-\fft{i}2\gamma\cdot\partial B_1
-\fft{i}2e^{A-2B_1+C}(1-|\alpha|^2)\sigma_3
+\fft{i}8e^{A-2B_1-2B_2-C}(4+\hat\phi_0)\nonumber\\
&&+\fft{i}4e^{-B_1+B_2}
\Bigl(\gamma\cdot D(\bar\alpha\sigma_+-\alpha\sigma_-)+e^{A-B_2}(2e^{C-B_2}
+3e^{B_2-C})(\bar\alpha\sigma_++\alpha\sigma_-)\Bigr)\nonumber\\
&&-\fft18e^{-2B_1-2B_2}\gamma\cdot\hat{\mathbb A}_1
+\fft1{16}e^{-A-C}\gamma\cdot(e^{-2B_1}\hat p_{21}-e^{-2B_2}\hat p_{22})
\sigma_3\biggr]\varepsilon\nonumber\\
&&-\fft{v_i}4\biggl[\fft1{24}(e^{-2A}\gamma\cdot\hat g_3^i
-3ie^{-A-C}\gamma\cdot\hat g_2^i)\nonumber\\
&&+\fft{i}{4}\left(\gamma\cdot(-3e^{-2B_1}\hat g_{11}^i+e^{-2B_2}\hat g_{12}^i)
-ie^{A-C}(-3e^{-2B_1}\hat j_{01}^i+e^{-2B_2}\hat j_{02}^i)\right)\sigma_3
\nonumber\\
&&+e^{-B_1-B_2}\left(\gamma\cdot(\hat f_1^i\sigma_+
-\hat{\bar f}_1^i\sigma_-)
-ie^{A-C}(\hat f_0^i\sigma_+-\hat{\bar f}_0^i\sigma_-)\right)\biggr]
\sigma_2\varepsilon^c,
\end{eqnarray}
\begin{eqnarray}
\delta\psi^{(7)}&=&\biggl[-\fft{i}2\gamma\cdot\partial B_2
-\fft{i}2e^{A-2B_2+C}\sigma_3
+\fft{i}8e^{A-2B_1-2B_2-C}(4+\hat\phi_0)\nonumber\\
&&-\fft{i}4e^{-B_1+B_2}
\Bigl(\gamma\cdot D(\bar\alpha\sigma_+-\alpha\sigma_-)-e^{A-B_2}(2e^{C-B_2}
-3e^{B_2-C})(\bar\alpha\sigma_++\alpha\sigma_-)\Bigr)\nonumber\\
&&-\fft18e^{-2B_1-2B_2}\gamma\cdot\hat{\mathbb A}_1
-\fft1{16}e^{-A-C}\gamma\cdot(e^{-2B_1}\hat p_{21}-e^{-2B_2}\hat p_{22})
\sigma_3\biggr]\varepsilon\nonumber\\
&&-\fft{v_i}4\biggl[\fft1{24}(e^{-2A}\gamma\cdot\hat g_3^i
-3ie^{-A-C}\gamma\cdot\hat g_2^i)\nonumber\\
&&+\fft{i}{4}\left(\gamma\cdot(e^{-2B_1}\hat g_{11}^i-3e^{-2B_2}\hat g_{12}^i)
-ie^{A-C}(e^{-2B_1}\hat j_{01}^i-3e^{-2B_2}\hat j_{02}^i)\right)\sigma_3
\nonumber\\
&&+e^{-B_1-B_2}\left(\gamma\cdot(\hat f_1^i\sigma_+-\hat{\bar f}_1^i\sigma_-)
-ie^{A-C}(\hat f_0^i\sigma_+-\hat{\bar f}_0^i\sigma_-)\right)\biggr]
\sigma_2\varepsilon^c,
\end{eqnarray}
\begin{eqnarray}
\delta\psi^{(9)}&=&\biggl[-\fft{i}2\gamma\cdot\partial C
+\fft{i}2(e^{A-2B_1+C}(1-|\alpha|^2)
+e^{A-2B_2+C}-3e^{A-C})\sigma_3\nonumber\\
&&+\fft{i}8e^{A-2B_1-2B_2-C}(4+\hat\phi_0)
-\fft{i}2e^{A-B_1}(2e^{C-B_2}-3e^{B_2-C})(\bar\alpha\sigma_++\alpha\sigma_-)
\nonumber\\
&&+\fft18e^{-2B_1-2B_2}\gamma\cdot\hat{\mathbb A}_1
+\fft18e^{C-A}\gamma\cdot F_2+\fft1{16}e^{-A-C}
\gamma\cdot(e^{-2B_1}\hat p_{21}+e^{-2B_2}\hat p_{22})\sigma_3\nonumber\\
&&+\fft{i}4e^{-A-B_1-B_2-C}\gamma\cdot(\hat q_2\sigma_+
-\hat{\bar q}_2\sigma_-)\biggr]\varepsilon\nonumber\\
&&-\fft{v_i}4\biggl[\fft1{24}(e^{-2A}\gamma\cdot\hat g_3^i
+9ie^{-A-C}\gamma\cdot\hat g_2^i)\nonumber\\
&&+\fft{i}{4}\left(\gamma\cdot(e^{-2B_1}\hat g_{11}^i+e^{-2B_2}\hat g_{12}^i)
+3ie^{A-C}(e^{-2B_1}\hat j_{01}^i+e^{-2B_2}\hat j_{02}^i)\right)\sigma_3
\nonumber\\
&&-e^{-B_1-B_2}\left(\gamma\cdot(\hat f_1^i\sigma_+-\hat{\bar f}_1^i\sigma_-)
+3ie^{A-C}(\hat f_0^i\sigma_+-\hat{\bar f}_0^i\sigma_-)\right)\biggr]
\sigma_2\varepsilon^c.
\end{eqnarray}
Finally, the shifted gravitino
\begin{equation}
\hat\psi_\alpha\equiv\psi_\alpha+\fft{i}3\gamma_\alpha(2\psi^{(5)}+2\psi^{(7)}+\psi^{(9)}),
\end{equation}
has the variation
\begin{eqnarray}\label{eq:N4gravsusy}
\delta\hat\psi_\alpha
&=&\biggl[\hat\nabla_\alpha-\fft{3i}2A_\alpha\sigma_3
+\fft{i}{4\tau_2}\partial_\alpha\tau_1-\fft{i}4e^{-2B_1-2B_2}
\hat{\mathbb A}_\alpha\nonumber\\
&&+\gamma_\alpha\Bigl(\fft16(e^{A-2B_1+C}(1-|\alpha|^2)+e^{A-2B_2+C}+3e^{A-C})
\sigma_3\nonumber\\
&&-\fft1{12}e^{A-2B_1-2B_2-C}(4+\hat\phi_0)\Bigr)
+\fft{i}{24}e^{C-A}(\gamma_\alpha{}^{\beta\gamma}
-4\delta_\alpha^\beta\gamma^\gamma)F_{\beta\gamma}\nonumber\\
&&-\fft{i}{24}e^{-A-C}(\gamma_\alpha{}^{\beta\gamma}
-4\delta_\alpha^\beta\gamma^\gamma)(e^{-2B_1}\hat p_{1\,\beta\gamma}
+e^{-2B_2}\hat p_{2\,\beta\gamma})\sigma_3\nonumber\\
&&-\fft12e^{-B_1+B_2}D_\alpha(\bar\alpha\sigma_+-\alpha\sigma_-)
-\fft16\gamma_\alpha e^{A-B_1}
(2e^{C-B_2}+3e^{B_2-C})(\bar\alpha\sigma_++\alpha\sigma_-)\nonumber\\
&&+\fft16e^{-A-B_1-B_2-C}(\gamma_\alpha{}^{\beta\gamma}
-4\delta_\alpha^\beta\gamma^\gamma)(\hat q_{\beta\gamma}\sigma_+
-\hat{\bar q}_{\beta\gamma}\sigma_-)\biggr]\varepsilon\nonumber\\
&&-iv_i\biggl[\fft1{36}e^{-2A}(\gamma_\alpha{}^{\beta\gamma\delta}
-\ft32\delta_\alpha^\beta\gamma^{\gamma\delta})\hat g_{\beta\gamma\delta}
-\fft{i}{24}e^{-A-C}(\gamma_\alpha{}^{\beta\gamma}
-4\delta_\alpha^\beta\gamma^\gamma)\hat g_{\beta\gamma}\nonumber\\
&&-\fft{i}4(e^{-2B_1}\hat g_{1\,\alpha}+e^{-2B_2}\hat g_{2\,\alpha})\sigma_3
-\fft1{12}e^{A-C}\gamma_\alpha(e^{-2B_1}\hat j_{01}+e^{-2B_2}\hat j_{02})
\sigma_3\nonumber\\
&&+e^{-B_1-B_2}(\hat f_\alpha\sigma_+-\hat{\bar f}_\alpha\sigma_-)
-\fft{i}3e^{A-B_1-B_2-C}\gamma_\alpha(\hat f_0\sigma_+-\hat{\bar f}_0\sigma_-)
\biggr]\sigma_2\varepsilon^c.\nonumber\\
\end{eqnarray}
The above constitute the most generic supersymmetry variations which are invariant under the $\mathrm{SU}(2)\times\mathrm{SU}(2)$ isometry group of $T^{1,1}.$ In section \ref{sec:N2} we will analyze two specific supersymmetric truncations of these variations. This entails truncating the bosonic field content and choosing an appropriate $\mathcal N = 2$ projection of the killing spinor $\epsilon.$ However, before doing so, and for the remainder of this section, we will keep the full $\mathcal N = 4$ field content and present the dimensional reduction of the fermion equations of motion.

\subsection{$D=5$, $\mathcal N = 4$ fermion equations of motion}
\label{sec:N4eom}

We now present the $\mathcal N = 4$ fermion equations of motion in the dimensionally reduced
theory.  Much of this reduction is rather tedious, and we only present the final results.  However, for
completeness, the reduction of the ten-dimensional covariant derivative is given in
appendix \ref{app:covderiv}. We begin with the reduction of the IIB dilatino equation in
(\ref{eq:iibeom}), again expressing the fermions as Sp(2) doublets:
\begin{eqnarray}\label{eq:N4dileom}
0&=&\gamma^\alpha \mathcal D_\alpha\lambda\nonumber\\
&& +\Bigl[ \ft{i}{8} \gamma\cdot
\mathcal F^{(\lambda)} -\ft14(4+\phi_0)e^{A-2B_1-2B_2-C}
- \ft12(3e^{A-C}+(1-|\alpha|^2)e^{A-2B_1+C} + e^{A-2B_2+C})\sigma_3 \nn \\
&&\kern3em -\ft12e^{-B_1+B_2}\gamma^\alpha D_\alpha(\bar\alpha\sigma_+ - \alpha\sigma_-)
+ (\ft32 e^{A-B_1+B_2-C} +e^{A-B_1-B_2+C})(\bar\alpha \sigma_+ + \alpha\sigma_-) \Big]
\lambda\nn \\
&& + v_i\Big[\ft19e^{-2A}\gamma\cdot\hat g_3^i+\ft{i}6e^{-A-C}\gamma\cdot\hat g_2^i
+\big(-ie^{-2B_1}\gamma^\alpha \hat g_{1\alpha} + \ft13e^{A-C}(4e^{-2B_1}\hat j^i_{01}
+e^{-2B_2}\hat j^i_{02})\big)\sigma_3 \nn \\
&&\kern3em+ 2e^{-B_1-B_2}\gamma^\alpha(\hat{f}^i_\alpha\sigma_+ - \hat{\bar{f}}^i_\alpha\sigma_-)
+ \ft{10i}3e^{A-B_1-B_2-C}(\hat{f}^i_0\sigma_+ - \hat{\bar{f}}^i_0\sigma_-) \Big]
\psi^{(5)}\nn \\
&&+ v_i\Big[\ft19e^{-2A}\gamma\cdot\hat g_3^i+\ft{i}6e^{-A-C}\gamma\cdot\hat g_2^i
+\big( - ie^{-2B_2}\gamma^\alpha \hat g_{2\alpha} + \ft13e^{A-C}(e^{-2B_1}\hat j^i_{01}
+4 e^{-2B_2}\hat j^i_{02})\big)\sigma_3 \nn \\
&&\kern3em+ 2e^{-B_1-B_2}\gamma^\alpha(\hat{f}^i_\alpha\sigma_+ - \hat{\bar{f}}^i_\alpha\sigma_-)
+ \ft{10i}3e^{A-B_1-B_2-C}(\hat{f}^i_0\sigma_+ - \hat{\bar{f}}^i_0\sigma_-) \Big]
\psi^{(7)}\nn\\
&& + v_i\Big[\ft1{18}e^{-2A}\gamma\cdot\hat g_3^i-\ft{i}6e^{-A-C}\gamma\cdot\hat g_2^i
+\ft23e^{A-C}(e^{-2B_1}\hat j^i_{01} + e^{-2B_2}\hat j^i_{02})\sigma_3\nn\\
&&\kern3em + \ft{8i}3e^{A-B_1-B_2-C}(\hat f^i_0\sigma_+ - \hat{\bar{f}}^i_0\sigma_-)\Big] \psi^{(9)}.
\end{eqnarray}
The five-dimensional supercovariant covariant derivative acts on the dilatino as
\begin{eqnarray}
\mathcal D_\alpha\lambda &=& \Big[\nabla_\alpha - \ft{3i}2 A_\alpha\sigma_3
+\ft{i}4e^{-2B_1-2B_2}\hat{\mathbb A}_\alpha +\ft{3i}{4\tau_2}\partial_\alpha\tau_1\Big]\lambda
-K(\lambda)\hat\psi_\alpha,
\end{eqnarray}
where the super-covariantization term, $K(\lambda)$, is the same operator which acts on $\epsilon$ in the dilatino variation (\ref{eq:susydil}).  We have also defined the field strength
\begin{equation}
\mathcal F^{(\lambda)} = e^{C-A}F_2
+ e^{-A-C}(e^{-2B_1}\hat p_{12}+e^{-2B_2}\hat p_{22})\sigma_3
+ 4ie^{-A-B_1-B_2-C}(\hat q_2\sigma_+ - \hat{\bar{q}}_2\sigma_-).
\end{equation}

Turning now to the reduction of the IIB gravitino equation of motion in (\ref{eq:iibeom}), let
\begin{equation}
E^A\equiv\Gamma^{ABC}\mathcal D_B\Psi_C+\ft{i}{48}v_i^*\Gamma\cdot F^{i*}\Gamma^A\lambda
-\ft{i}{4\tau_2}\Gamma\cdot\partial\tau\Gamma^A\lambda^c.
\label{eq:Psieom}
\end{equation}
Breaking this up into components, $\{E^\alpha,E^a,E^{\hat a},E^9\}$, we obtain separated
spin-1/2 equations of motion
\begin{eqnarray}
E^{(5)}&=&\Gamma^\alpha E^\alpha-3\Gamma^aE^a+\Gamma^{\hat a}E^{\hat a}+\Gamma^9E^9,
\nn\\
E^{(7)}&=&\Gamma^\alpha E^\alpha+\Gamma^aE^a-3\Gamma^{\hat a}E^{\hat a}+\Gamma^9E^9,
\nn\\
E^{(9)}&=&\Gamma^\alpha E^\alpha+\Gamma^aE^a+\Gamma^{\hat a}E^{\hat a}-7\Gamma^9E^9.
\end{eqnarray}
In addition, the five-dimensional gravitino equation of motion is simply
\begin{equation}
E^\alpha = 0.
\end{equation}
For the spin-1/2 equations of motion, we arrive at the following for $E^{(5)},$ $E^{(7)}$ and $E^{(9)},$ respectively:

\begin{eqnarray}\label{eq:psi5eom}
&&0= \gamma^\alpha \mathcal D_\alpha \psi^{(5)} \nn \\
&&+\Bigl[\ft{i}8\gamma\cdot \mathcal{F}^{(5)} -\ft5{12}e^{A-2B_1-2B_2-C}(4+\hat\phi_0) -\ft32\sigma_3e^{A-C} +\ft{13}6\sigma_3e^{A-2B_1+C}(1-|\alpha|^2) \nn \\
&&\kern2em-\ft12\sigma_3e^{A-2B_2+C}+\ft23 e^{A-B_1}
(e^{-B_2+C}+\ft32e^{B_2-C})(\bar\alpha\sigma_+ + \alpha\sigma_-) \nn\\
&&\kern2em-\ft{i}{12}v_i\bigl(-\ft16e^{-2A}\gamma\cdot\hat g_3^i-ie^{-A-C}\gamma\cdot\hat g_2^i
+6i\sigma_3e^{-2B_1}\gamma\cdot\hat g_{11}^i-3i\sigma_3e^{-2B_2}\gamma\cdot\hat g_{12}^i\nn\\
&&\kern3em+6e^{-B_1-B_2}\gamma\cdot(\hat f_1^i\sigma_+
-\hat{\bar f}_1^i\sigma_-)
-9\sigma_3e^{A-2B_1-C}\hat j_{01}^i+4\sigma_3e^{A-2B_2-C}\hat j_{02}^i\nn\\
&&\kern3em
+2ie^{A-B_1-B_2-C}(\hat f_0^i\sigma_-
+\hat{\bar f}_0^i\sigma_-)\bigr)\sigma_2\mathcal C\Bigr]\psi^{(5)}\nn\\
&&+\Bigl[\ft{i}2e^{-2B_1-2B_2}\gamma\cdot\hat{\Bbb A}-\ft12e^{-B_1+B_2}
\gamma\cdot D(\bar\alpha\sigma_+ - \alpha\sigma_-) -\ft{i}{6}\sigma_3e^{-A-C}\gamma\cdot(\ft12e^{-2B_1}\hat p_{21}+e^{-2B_2}\hat p_{22})\nn\\
&& \kern2em-\ft23e^{A-2B_1-2B_2-C}(4+\hat\phi_0) +\ft23\sigma_3e^{A-2B_1+C}(1-|\alpha|^2) \nn \\
&&\kern2em -e^{A-B_1}(-\ft53e^{-B_2+C}+\ft12e^{B_2-C})
(\bar\alpha\sigma_+ + \alpha\sigma_-) -\ft{i}{12}v_i\bigl(\ft13e^{-2A}\gamma\cdot\hat g_3^i+\ft{i}2e^{-A-C}\gamma\cdot\hat g_2^i\nn\\
&&\kern3em-3i\sigma_3e^{-2B_2}\gamma\cdot\hat g_{12}^i
-3\sigma_3e^{A-2B_1-C}\hat j_{01}^i+4\sigma_3e^{A-2B_2-C}\hat j_{02}^i\nn\\
&&\kern3em-6e^{-B_1-B_2}\gamma\cdot(\hat f_1^i\sigma_+
-\hat{\bar f}_1^i\sigma_-)
-10ie^{A-B_1-B_2-C}(\hat f_0^i\sigma_+
-\hat{\bar f}_0^i\sigma_-)\bigr)\sigma_2\mathcal C\Bigr]\psi^{(7)}\nn\\
&&+\Bigl[\ft{i}{12}\sigma_3e^{-A-C}\gamma\cdot(e^{-2B_1}\hat p_{21}-e^{-2B_2}\hat p_{22})-\ft13e^{A-2B_1-2B_2-C}(4+\hat\phi_0)-\ft23\sigma_3e^{A-2B_1+C}(1-|\alpha|^2)\nn\\
&&\kern2em-e^{A-B_1}(\ft23e^{-B_2+C}-2e^{B_2-C})
(\bar\alpha\sigma_+ + \alpha\sigma_-)-\ft{i}{12}v_i\bigl(\ft16e^{-2A}\gamma\cdot\hat g_3^i-\ft{i}2e^{-A-C}\gamma\cdot\hat g_2^i\nn\\
&&
\kern3em-6\sigma_3e^{A-2B_1-C}\hat j_{01}^i+2\sigma_3e^{A-2B_2-C}\hat j_{02}^i-8ie^{A-B_1-B_2-C}(\hat f_0^i\sigma_+
-\hat{\bar f}_0^i\sigma_-)\bigr)\sigma_2\mathcal C\Bigr]\psi^{(9)} \nn\\
&&+\Bigl[\ft{1}{16}v_i^*\bigl(\ft16e^{-2A}\gamma\cdot\hat g_3^i
+\ft{i}2e^{-A-C}\gamma\cdot\hat g_2^i-3i\sigma_3e^{-2B_1}\gamma\cdot\hat g_{11}^i
+i\sigma_3e^{-2B_2}\gamma\cdot\hat g_{12}^i\nn\\
&&\kern3em+3\sigma_3e^{A-2B_1-C}\hat j_{01}^i
-\sigma_3e^{A-2B_2-C}\hat j_{02}^i\nn\\
&&\kern3em+4e^{-B_1-B_2}\gamma\cdot(\hat f_1^i\sigma_+-\tilde{\bar f}_1^i\sigma_-)
+4ie^{A-B_1-B_2-C}(\hat f_0^i\sigma_+-\hat{\bar f}_0^i\sigma_-)\bigr)\Bigr]\lambda,
\end{eqnarray}

\begin{eqnarray}\label{eq:psi7eom}
&&0=\gamma^\alpha\mathcal D_\alpha\psi^{(7)} \nn \\
&&+\Bigl[\ft{i}2e^{-2B_1-2B_2}\gamma\cdot\hat{\Bbb A}-\ft12e^{-B_1+B_2}
\gamma\cdot D(\bar\alpha\sigma_+ - \alpha\sigma_-)  -\ft{i}6\sigma_3e^{-A-C}\gamma\cdot(e^{-2B_1}\hat p_{21}+\ft12e^{-2B_2}\hat p_{22})\nn\\
&&\kern3em -\ft23e^{A-2B_1-2B_2-C}(4+\hat\phi_0) + \ft23\sigma_3e^{A-2B_2+C}-e^{A-B_1}(-\ft53e^{-B_2+C}+\ft52e^{B_2-C})(\bar\alpha\sigma_+ + \alpha\sigma_-)\nn\\
&&\kern2em-\ft{i}{12}v_i\bigl(\ft13e^{-2A}\gamma\cdot\hat g_3^i+\ft{i}2e^{-A-C}\gamma\cdot\hat g_2^i
-3i\sigma_3e^{-2B_1}\gamma\cdot\hat g_{11}^i
+4\sigma_3e^{A-2B_1-C}\hat j_{01}^i\nn\\
&&\kern3em-3\sigma_3e^{A-2B_2-C}\hat j_{02}^i-6e^{-B_1-B_2}\gamma\cdot(\hat f_1^i\sigma_+
-\hat{\bar f}_1^i\sigma_-)\nn\\
&&\kern3em -10ie^{A-B_1-B_2-C}(\hat f_0^i\sigma_+ -\hat{\bar f}_0^i\sigma_-)\bigr)\sigma_2\mathcal C\Bigr]\psi^{(5)}\nn\\
&&+\Bigl[\ft{i}8\gamma\cdot\mathcal{F}^{(7)}-\ft5{12}e^{A-2B_1-2B_2-C}(4+\hat\phi_0)-\ft32\sigma_3e^{A-C}-\ft12\sigma_3e^{A-2B_1+C}(1-|\alpha|^2)\nn\\
&&\kern3em+\ft{13}6\sigma_3e^{A-2B_2+C}+\ft23e^{A-B_1}
(e^{-B_2+C}-\ft32e^{B_2-C})(\bar\alpha\sigma_+ + \alpha\sigma_-)\nn \\
&&\kern2em -\ft{i}{12}v_i\bigl(-\ft16e^{-2A}\gamma\cdot\hat g_3^i-ie^{-A-C}\gamma\cdot\hat g_2^i
-3i\sigma_3e^{-2B_1}\gamma\cdot\hat g_{11}^i\nn\\
&&\kern3em+6i\sigma_3e^{-2B_2}\gamma\cdot\hat g_{12}^i
+4\sigma_3e^{A-2B_1-C}\hat j_{01}^i-9\sigma_3e^{A-2B_2-C}\hat j_{02}^i\nn\\
&&\kern3em+6e^{-B_1-B_2}\gamma\cdot(\hat f_1^i\sigma_+
-\hat{\bar f}_1^i\sigma_-)
+2ie^{A-B_1-B_2-C}(\hat f_0^i\sigma_+
-\hat{\bar f}_0^i\sigma_-)\bigr)\sigma_2\mathcal C\Bigr]\psi^{(7)}\nn\\
&&+\Bigl[-\ft{i}{12}\sigma_3e^{-A-C}\gamma\cdot(e^{-2B_1}\hat p_{21}
-e^{-2B_2}\hat p_{22}) -\ft13e^{A-2B_1-2B_2-C}(4+\hat\phi_0) -\ft23e^{A-2B_2+C} \nn \\
&&\kern2em -e^{A-B_1}(\ft23e^{-B_2+C}+2e^{B_2-C})
(\bar\alpha\sigma_+ + \alpha\sigma_-) -\ft{i}{12}v_i\bigl(\ft16e^{-2A}\gamma\cdot\hat g_3^i-\ft{i}2e^{-A-C}\gamma\cdot\hat g_2^i \nn \\
&& \kern3em +2\sigma_3e^{A-2B_1-C}\hat j_{01}^i-6\sigma_3e^{A-2B_2-C}\hat j_{02}^i -8ie^{A-B_1-B_2-C}(\hat f_0^i\sigma_+
-\hat{\bar f}_0^i\sigma_-)\bigr)\sigma_2 \mathcal C \Bigr]\psi^{(9)},\nn\\
&&+\Bigl[\ft{1}{16}v_i^*\bigl(\ft16e^{-2A}\gamma\cdot\hat g_3^i
+\ft{i}2e^{-A-C}\gamma\cdot\hat g_2^i+i\sigma_3e^{-2B_1}\gamma\cdot\hat g_{11}^i
-3i\sigma_3e^{-2B_2}\gamma\cdot\hat g_{12}^i\nn\\
&&\kern3em-\sigma_3e^{A-2B_1-C}\hat j_{01}^i
+3\sigma_3e^{A-2B_2-C}\hat j_{02}^i\nn\\
&&\kern3em+4e^{-B_1-B_2}\gamma\cdot(\hat f_1^i\sigma_+-\tilde{\bar f}_1^i\sigma_-)
+4ie^{A-B_1-B_2-C}(\hat f_0^i\sigma_+-\hat{\bar f}_0^i\sigma_-)\bigr)\Bigr]\lambda,
\end{eqnarray}

\begin{eqnarray}\label{eq:psi9eom}
&&0=\gamma^\alpha\mathcal D_\alpha\psi^{(9)} \nn \\
&&+\Big[-\ft{i}2e^{-2B_1-2B_2}\gamma\cdot\hat{\Bbb A}-\ft{i}{6}e^{-A+C}\gamma\cdot F +\ft{i}6\sigma_3e^{-A-C}\gamma\cdot(e^{-2B_1}\hat p_{21}-\ft12e^{-2B_2}\hat p_{22})\nn \\
&&\kern3em -\ft1{6}e^{-A-B_1-B_2-C}\gamma\cdot(\hat q_2\sigma_+
-\hat{\bar q}_2\sigma_-) -\ft23e^{A-2B_1-2B_2-C}(4+\hat\phi_0) + 2\sigma_3e^{A-C}\nn \\
&&\kern3em-\ft83\sigma_3e^{A-2B_1+C}(1-|\alpha|^2)-\ft23\sigma_3e^{A-2B_2+C} \nn\\
&&\kern3em -\ft{10}3e^{A-B_1}(e^{-B_2+C}-\ft32e^{B_2-C})
(\bar\alpha\sigma_+ +\alpha\sigma_-) -\ft{i}{12}v_i\bigl(\ft13e^{-2A}\gamma\cdot\hat g_3^i-\ft{3i}2e^{-A-C}\gamma\cdot\hat g_2^i  \nn\\
&&\kern3em
-3i\sigma_3e^{-2B_1}\gamma\cdot\hat g_{11}^i \nn -12\sigma_3e^{A-2B_1-C}\hat j_{01}^i-3\sigma_3e^{A-2B_2-C}\hat j_{02}^i\nn\\
&&\kern3em+6e^{-B_1-B_2}\gamma\cdot(\hat f_1^i\sigma_+
-\hat{\bar f}_1^i\sigma_-)
-30ie^{A-B_1-B_2-C}(\hat f_0^i\sigma_+
-\hat{\bar f}_0^i\sigma_-)\bigr)\sigma_2\mathcal C \Bigr]\psi^{(5)}\nn\\
&&+\Bigl[-\ft{i}2e^{-2B_1-2B_2}\gamma\cdot\hat{\Bbb A}-\ft{i}{6}e^{-A+C}\gamma\cdot F
-\ft{i}{6}\sigma_3e^{-A-C}\gamma\cdot(\ft12e^{-2B_1}\hat p_{21}-e^{-2B_2}\hat p_{22})\nn\\
&&\kern3em -\ft1{6}e^{-A-B_1-B_2-C}\gamma\cdot(\hat q_2\sigma_+
-\hat{\bar q}_2\sigma_-) -\ft23e^{-2B_1-2B_2-C}(4+\hat\phi_0) +2\sigma_3e^{A-C} \nn \\
&& \kern3em -\ft23\sigma_3e^{A-2B_1+C}(1-|\alpha|^2) -\ft83\sigma_3e^{A-2B_2+C}\nn\\
&&\kern3em-e^{A-B_1}(\ft{10}3e^{-B_2+C}+e^{B_2-C})
(\bar\alpha\sigma_++\alpha\sigma_-) - \ft{i}{12}v_i\bigl(\ft13e^{-2A}\gamma\cdot\hat g_3^i-\ft{3i}2e^{-A-C}\gamma\cdot\hat g_2^i \nn \\
&& \kern3em
-3i\sigma_3e^{-2B_2}\gamma\cdot\hat g_{12}^i
-3\sigma_3e^{A-2B_1-C}\hat j_{01}^i-12\sigma_3e^{A-2B_2-C}\hat j_{02}^i \nn\\
&&\kern3em+6e^{-B_1-B_2}\gamma\cdot(\hat f_1^i\sigma_+
-\hat{\bar f}_1^i\sigma_-)
-30ie^{A-B_1-B_2-C}(\hat f_0^i\sigma_+
-\hat{\bar f}_0^i\sigma_-)\bigr)\sigma_2\mathcal C  \Bigr]\psi^{(7)}\nn\\
&&+\Bigl[\ft{i}8\gamma\cdot\mathcal{F}^{(9)}-\ft12e^{-B_1+B_2}
\gamma\cdot D(\bar\alpha\sigma_+-\alpha\sigma_-) -\ft1{12}e^{A-2B_1-2B_2-C}(4+\hat\phi_0) +\ft52\sigma_3e^{A-C} \nn \\
&&\kern3em+\ft16\sigma_3e^{A-2B_1+C}(1-|\alpha|^2)+\ft16\sigma_3e^{A-2B_2+C}\nn\\
&&\kern3em +e^{A-B_1}(\ft13e^{-B_2+C}+\ft52e^{B_2-C})
(\bar\alpha\sigma_++\alpha\sigma_-) - \ft{i}{12}v_i\bigl(-\ft13e^{-2A}\gamma\cdot\hat g_3^i \nn\\
&&\kern3em -3i\sigma_3\gamma\cdot(e^{-2B_1}\hat g_{11}^i+e^{-2B_2}\hat g_{12}^i) +3\sigma_3e^{A-C}(e^{-2B_1}\hat j_{01}^i+e^{-2B_1}\hat j_{02}^i)\nn\\
&&\kern3em+12e^{-B_1-B_2}\gamma\cdot(\hat f_1^i\sigma_+
-\hat{\bar f}_1^i\sigma_-)
-12i(\hat f_0^i\sigma_+
-\hat{\bar f}_0^i\sigma_-)\bigr)\sigma_2 \mathcal C\Bigr]\psi^{(9)} \nn \\
&&+\Bigl[\ft{1}{16}v_i^*\bigl(\ft16e^{-2A}\gamma\cdot\hat g_3^i
-\ft{3i}2e^{-A-C}\gamma\cdot\hat g_2^i+i\sigma_3e^{-2B_1}\gamma\cdot\hat g_{11}^i
+i\sigma_3e^{-2B_2}\gamma\cdot\hat g_{12}^i\nn\\
&&\kern3em+3\sigma_3e^{A-2B_1-C}\hat j_{01}^i
+3\sigma_3e^{A-2B_2-C}\hat j_{02}^i\nn\\
&&\kern3em-4e^{-B_1-B_2}\gamma\cdot(\hat f_1^i\sigma_+
-\tilde{\bar f}_1^i
\sigma_-)+12ie^{A-B_1-B_2-C}(\hat f_0^i\sigma_+-\hat{\bar f}_0^i\sigma_-)\bigr)\Bigr]\lambda.
\end{eqnarray}
Here, $\mathcal C$ denotes the five-dimensional charge conjugation operator: $\mathcal C\psi
=\psi^c$.  Similarly to the dilatino, the five-dimensional supercovariant derivative acts on the
spin-1/2 components of the IIB gravitino as
\begin{eqnarray}
\mathcal D_\alpha\psi^{(i)} &=& \Big[\nabla_\alpha - \ft{3i}2 A_\alpha\sigma_3
+\ft{i}4\eta^{(i)}e^{-2B_1-2B_2}\mathbb A_\alpha +\ft{i}{4\tau_2}\partial_\alpha\tau_1\Big]\psi^{(i)}
-K(\psi^{(i)})\hat\psi_\alpha,
\end{eqnarray}
where $\eta^{(5)}=\eta^{(7)}=+1$ and $\eta^{(9)}=-1$.  We have also defined the following
combinations of field strengths
\begin{eqnarray}
\mathcal{F}^{(5)}&=&e^{C-A} F_2 +\ft13e^{-A-C}(e^{-2B_1}\hat p_{21}
-e^{-2B_2}\hat p_{22})\sigma_3 - 4ie^{-A-B_1-B_2-C}(\hat q_2\sigma_+
-\hat{\bar q}_2\sigma_-), \nn \\
\mathcal{F}^{(7)}&=&e^{C-A} F_2 -\ft13e^{-A-C}(e^{-2B_1}\hat p_{21}-e^{-2B_2}\hat p_{22})\sigma_3
-4ie^{-A-B_1-B_2-C}(\hat q_2\sigma_+ -\hat{\bar q}_2\sigma_-), \nn \\
\mathcal{F}^{(9)}&=& -\ft53e^{C-A} F_2 -\ft13e^{-A-C}(e^{-2B_1}\hat p_{21}+e^{-2B_2}\hat p_{22})\sigma_3
-\ft{4i}{3}e^{-A-B_1-B_2-C}\gamma\cdot(\hat q_2\sigma_+
-\hat{\bar q}_2\sigma_-). \nn \\
\end{eqnarray}

Finally, the gravitino equation arranges itself into terms appropriate for supercovariantization,
\begin{eqnarray}\label{eq:N4graveom}
0&=& \gamma^{\alpha\beta\gamma}\mathcal D_\beta\hat{\psi}_\gamma - \ft23 \tilde{K}(\psi^{(5)})\gamma^\alpha(5\psi^{(5)}+2\psi^{(7)}+\psi^{(9)}) - \ft23 \tilde{K}(\psi^{(7)})\gamma^\alpha(2\psi^{(5)}+5\psi^{(7)}+\psi^{(9)})\nn \\
&& - \ft23 \tilde{K}(\psi^{(9)})\gamma^\alpha(\psi^{(5)}+\psi^{(7)}+2\psi^{(9)}) - \ft12 \tilde{K}(\lambda) \gamma^\alpha\lambda.
\end{eqnarray}

In this section we have presented the fermionic sector of type IIB supergravity reduced on $T^{1,1}$ while allowing for the most general deformations which preserve the isometries of $T^{1,1}$. This yields a five-dimensional $\mathcal N = 4$ supergravity theory described by the supersymmetry transformations and equations of motion detailed above. In the following section we will present and analyze the supersymmetry variations of two particular $\mathcal N = 2$ truncations of this system. It would be interesting to also consider the corresponding $\mathcal N = 2$ equations of motion presented in this section.  While we do not do this, it should be straightforward, although tedious, to do so using the results in the following section.

\section{$\mathcal N = 2$ truncations}
\label{sec:N2}

We are interested in the truncation to $\mathcal N = 2$ supergravity which involves setting the massive gravitino multiplet to zero. In \cite{Cassani:2010na} it is shown that, at the level of the bosonic equations of motion, this is consistent only if we also break the so-called $\mathcal N = 4$ Betti vector multiplet, which in terms of the bosonic content consists of
\begin{equation}
\{r_2, \, (B_1-B_2), \, e_0^i, \, \alpha\}.
\end{equation}
The truncations presented in \cite{Cassani:2010na} keep either a massless $\mathcal N = 2$ vector multiplet%
\footnote{Note that $r_2$ represents the field strength of this massless vector.}
$\{r_2, \, (B_1-B_2)\}$,
or an $\mathcal N = 2$ hyper-multiplet $\{e_0^i,\, \alpha\}$, termed the Betti vector multiplet and Betti hyper-multiplet, respectively. Note that these truncations represent two explicit truncations of the $\mathcal N = 4$ theory to $\mathcal N = 2$ and that other truncations probably exist for different choices of the spacetime Killing spinor.

\subsection{Truncating out the Betti hyper-multiplet}

As mentioned, the $\mathcal N = 2$ truncations which we will consider include, on top of the universal field content in all $SE_5$ reductions detailed in \cite{Cassani:2010uw,Gauntlett:2010vu,Liu:2010sa}, either the Betti vector multiplet or the Betti hyper-multiplet \cite{Cassani:2010na}. We begin by presenting the truncation keeping the Betti vector multiplet, which is the more straightforward of the two, and will discuss the Betti hyper multiplet in the following section.

The truncation proceeds by setting the massive gravitino multiplet modes to zero
\begin{equation}
\{b_2^i,\, b_1^i,\, q_2,\, c_0^i \} = 0,
\end{equation}
along with the fields in the Betti hyper-multiplet,
\begin{equation}
\{\alpha,\, e_0^i,\, j_0^i\} = 0.
\end{equation}

Since we are reducing to $\mathcal N = 2$ supergravity, we must choose an appropriate embedding of $\mathcal N = 2$ within $\mathcal N=4$.  In terms of the supersymmetry parameter, $\varepsilon$, this turns out to be
\begin{equation}\label{eq:bvecspin}
\varepsilon = \btop{\varepsilon_1}{0}.
\end{equation}
With this choice we find that the variations of the modes corresponding to the massive gravitino multiplet all vanish identically, so that it is consistent to set those fermions to zero,
\begin{equation}\label{eq:mgtinofermtrunc}
\{\hat\psi_{2\,\alpha}, \, \psi^{(5)}_2, \, \psi^{(7)}_2, \, \psi^{(9)}_2, \, \lambda_1 \} = 0,
\end{equation}
leaving the upper components of the fields from the 10-dimensional gravitino and the lower component of the dilatino remaining.

To illustrate the multiplet structure, we arrange the supersymmetry variations into appropriate linear combinations,
\begin{eqnarray}
&&\psi^{m=11/2} = 2(\psi_1^{(5)} + \psi_1^{(7)}) + \psi_1^{(9)}, \nonumber \\
&&\psi^{m=-9/2} = \ft12(\psi_1^{(5)}+\psi_1^{(7)}) - \psi_1^{(9)},\nonumber \\
&&\psi^{m=-1/2} = \psi_1^{(5)} - \psi_1^{(7)},
\end{eqnarray}
which are based on the Kaluza-Klein mass eigenstates for the fermions in the
AdS$_5$ vacuum.  The spin-1/2 supersymmetry variations are then given by
\begin{eqnarray}\label{eq:BVsusy1}
\delta\lambda&=&\fft{1}{2\tau_2}\gamma\cdot\partial\tau \varepsilon^c-ie^{-B_1-B_2}v_i\left(\gamma\cdot\bar f_1^i
-ie^{A-C}\bar f_0^i\right)\varepsilon, \nonumber \\
\delta\psi^{m=11/2}&=&\Bigl[-i\gamma\cdot\partial (B_1 +B_2 + \ft12C)
-\ft{i}2(e^{A-2B_1+C} +e^{A-2B_2+C} +3e^{A-C})\nonumber\\
&&-\ft38e^{-2B_1-2B_2}\gamma\cdot{\mathbb A}_1
+ \ft18e^{C-A}\gamma\cdot F_2 + \ft1{16}e^{-A-C}\gamma\cdot(e^{-2B_1}p_{21}+e^{-2B_2}p_{22}) \nonumber \\
&&+\ft{5i}8e^{A-2B_1-2B_2-C}(4+\phi_0)\Bigr]\varepsilon - v_ie^{-B_1-B_2}\left(\ft{3i}4\gamma\cdot f_1^i
+ \ft74e^{A-C}f_0^i\right)\varepsilon^c, \nonumber\\
\delta\psi^{m=-9/2}&=&\Bigl[-\ft{i}2\gamma\cdot\partial(\ft12(B_1+B_2) - C)
-\ft{3i}4(e^{A-2B_1+C} + e^{A-2B_2+C} - 2e^{A-C})\nonumber\\
&&-\ft14e^{-2B_1-2B_2}\gamma\cdot{\mathbb A}_1 - \ft18e^{C-A}\gamma\cdot F_2
-\ft1{16}e^{-A-C}\gamma\cdot(e^{-2B_1}p_{21}+e^{-2B_2} p_{22})
\Bigr]\varepsilon\nonumber\\
&&-v_ie^{-B_1-B_2}\left(\ft{i}2\gamma\cdot f_1^i
- \ft12e^{A-C} f_0^i\right)
\varepsilon^c, \nonumber \\
\delta\psi^{m=-1/2}&=&\Bigl[-\ft{i}2\gamma\cdot\partial (B_1-B_2)
-\ft{i}2(e^{A-2B_1+C}
-e^{A-2B_2+C}) \nonumber \\
&&+\ft1{8}e^{-A-C}\gamma\cdot(e^{-2B_1}p_{21}-e^{-2B_2}p_{22})\Bigr]\varepsilon,
\end{eqnarray}
and the shifted gravitino variation is
\begin{eqnarray}\label{eq:BVsusy2}
\delta\hat\psi_\alpha
&=&\Bigl[D_\alpha +\ft{i}{24}e^{C-A}(\gamma_\alpha{}^{\beta\gamma}
-4\delta_\alpha^\beta\gamma^\gamma)(F_{\beta\gamma} - e^{-2B_1-2C}p_{1\,\beta\gamma}
- e^{-2B_2-2C}p_{2\,\beta\gamma}) \nonumber \\
&&+\ft1{6}\gamma_\alpha\Bigl(e^{A-2B_1+C}+e^{A-2B_2+C}+3e^{A-C}-\ft1{2}e^{A-2B_1-2B_2-C}(4+\phi_0)\Bigr) \Bigr]\varepsilon \nonumber\\
&&+v_i\Bigl[e^{-B_1-B_2} f^i_\alpha -\ft{i}3e^{A-B_1-B_2-C}\gamma_\alpha f^i_0
\Bigr]\varepsilon^c,\nonumber\\
\end{eqnarray}
where $D_\alpha \equiv \nabla_\alpha - \ft{3i}2(A_\alpha
+\ft16e^{-2B_1-2B_2}\mathbb A_\alpha) + \ft{i}{4\tau_2}\partial_\alpha\tau_1$
is the fully gauge and gravitationally covariant derivative.

By setting $B_1 = B_2 = B$ and $p_{21} = p_{22} = p_2$ in the supersymmetry
transformations, we see that $\delta\psi^{m=-1/2}$ vanishes identically,
and we recover the results in \cite{Liu:2010pq}%
\footnote{Note that the conventions for the charge conjugate spinor
$\epsilon^c$ in five-dimensions is different in \cite{Liu:2010pq}.
Here it is defined as $\epsilon^c = \gamma_0 C_{4,1}\epsilon^*$, which
differs by a minus sign to those in \cite{Liu:2010pq}}.
It is thus apparent that this reduction on $T^{1,1}$ has an additional
fermionic mode $\psi^{m=-1/2}$ compared with the generic case. Along with
the bosonic fields $\{r_2, B_1-B_2\}$, the additional matter fill out a
massless $\mathcal N = 2$ vector multiplet,
\begin{equation}
\mathcal D(2,0,0)_0=D(3,\ft12,\ft12)_0+D(2\ft12,\ft12,0)_{-1}
+D(2\ft12,0,\ft12)_1+D(2,0,0)_0,
\end{equation}
which has previously been identified as the Betti vector multiplet.

While the Betti vector is a feature of the $T^{1,1}$ reduction, and is
absent in general, it is nevertheless curious to note that the spectrum of
Kaluza-Klein modes on $S^5$ contains these fields, namely a massless vector, an $m^2 = - 4$ scalar, and an $m=-1/2$ fermion, all at the lowest rungs of their
respective KK towers \cite{Kim:1985ez}%
\footnote{It should be noted that the only bottom rung with a massless vector
is filled by the $\mathcal N=2$ graviphoton.  However, this does not preclude
the identification of a second massless vector at the same massless level.}.
In the maximally supersymmetric $S^5$ case, the additional scalar lies in
the $\mathbf{20'}$ representation of SU(4).  Since this does not contain
any SU(3) singlets under $\rm SU(4)\supset SU(3)\times U(1)$, such a scalar
is absent in the squashed $S^5$ reduction as well as in the generic $SE_5$
reductions.  Nevertheless, as the Betti vector states lie at the bottom of
their respective KK-towers \cite{Ceresole:1999zs,Ceresole:1999ht}, consistency
of the $T^{1,1}$ truncation is obtained by restricting to singlets under
the $\rm SU(2)\times SU(2)$ isometry of $T^{1,1}$.

{}From the point of view of $\mathcal N=8$ supergravity truncated to
$\mathcal N=2$, this suggests a breaking of the $R$-symmetry according to
$\rm SU(4)\supset SU(2)\times SU(2)\times U(1)$ followed by a truncation to
singlets under both SU(2)'s.  However, even if such a truncation were
consistent, it would require keeping an infinite set of modes along the KK
towers, as representations of SU(4) containing singlets under
$\rm SU(2)\times SU(2)$ are seemingly too easy to come by. This is related to
the fact that a consistent truncation which keeps only a finite number of
modes results from isolating singlets under a transitively acting subgroup
of the isometry group, which is the case for $\rm SU(3) \subset SU(4)$ but is
not true for the standard embedding of $\rm SU(2)\times SU(2) \subset SU(4)$.

\subsubsection{$\mathcal N = 2$ superpotential}

One advantage of having worked out the supersymmetry variations is that it allows us to determine an appropriate $\mathcal N = 2$ superpotential for this truncation. From the gravitino variation written above we can read off the following superpotential
\begin{equation}
W = -\fft12(4+\phi_0)e^{A-2B_1-2B_2-C} + e^{A-2B_1+C} + e^{A-2B_1+C} + 3e^{A-C}.
\end{equation}
This superpotential reproduces the scalar potential via the relation
\begin{equation}
V = 2 (\mathcal G^{-1})^{ij} \partial_iW\partial_jW - \fft43 W^2,
\label{eq:potspot}
\end{equation}
where for this truncation the inverse scalar metric is given by
\begin{eqnarray}
(\mathcal G_{\{B_1,B_2,C\}}^{-1})^{ij} = \fft1{16}\begin{pmatrix}3&-1&-1\cr-1&3&-1\cr-1&-1&7
\end{pmatrix}, \,\,\, (\mathcal G_{\{b_0^1,b_0^2\}}^{-1})^{ij} =
\fft{1}{4\tau_2}e^{2B_1+2B_2}\begin{pmatrix}1&\tau_1\cr\tau_1&|\tau|^2
\end{pmatrix}, \,\,\, \mathcal G_{\tau}^{-1} = \tau_2^2.
\end{eqnarray}
A further truncation of the Betti vector, obtained by setting $B_1=B_2=B$,
then reproduces the superpotential obtained in \cite{Liu:2000gk,Liu:2010pq}
\begin{equation}
W = -\fft12(4+\phi_0)e^{A-4B-C} + 2e^{A-2B+C} + 3e^{A-C}.
\end{equation}

\subsection{Truncating out the Betti vector multiplet}\label{sec:bhyper}

The truncation which retains the Betti hyper-multiplet is a bit more interesting. We again begin by setting the massive gravitino modes to zero. To appropriately truncate the $\mathcal N = 2$ Betti vector multiplet we are required to set
\begin{equation}
r_2=0, \qquad e^{2(B_1-B_2)}=1-|\alpha|^2.
\end{equation}

It is natural to attempt to choose a simple embedding of $\mathcal N = 2$
supersymmetry in $\mathcal N=4$ as in (\ref{eq:bvecspin}). However, if we
did this, it would soon become apparent that we cannot truncate out the
fermions in the massive gravitino multiplet as in (\ref{eq:mgtinofermtrunc}).
So apparently there is some subtlety with this truncation. In particular,
since $\alpha$ encodes essentially a twisting of the internal vielbein, this twisting also affects the fermions, which take values in the spin bundle.
We find that an appropriate embedding requires that the supersymmetry
parameter $\varepsilon$ as well as all the fermions take the twisted form
\begin{equation}
\varepsilon = e^{\fft{i}{2}\vec\beta\cdot\vec\sigma}
\btop{\tilde\varepsilon_1}0,
\label{eq:twist}
\end{equation}
where
\begin{equation}
\vec\beta = \fft{k}{|\alpha|}(-\alpha_2,\alpha_1,0), \qquad k = \arctan\biggl(\fft{|\alpha|}{\sqrt{1-|\alpha|^2}}\biggr).
\end{equation}
Note that the un-rotated spinors will be denoted with a tilde.
This rotation is in retrospect expected based on the form of the
Killing spinors in general conifold reductions as discussed, {\it e.g.},
in \cite{Arean:2006nc}.

With the embedding (\ref{eq:twist}), we verify that it is consistent
to set the following fermions in the massive gravitino multiplet to zero,
\begin{equation}
\{\tilde\psi_{2\,\alpha}, \, (\tilde\psi^{(5)}_2 + \tilde\psi^{(7)}_2), \, \tilde\psi^{(9)}_2, \, \tilde\lambda_1 \} = 0,
\end{equation}
as they all have vanishing variation. Notice that we have not set
the difference, $\tilde\psi^{(5)}_2 - \tilde\psi^{(7)}_2$ to
zero. In order to have consistent supersymmetry variations it is necessary to make the identification
\begin{equation}
\tilde\psi^{(5)}_1 - \tilde\psi^{(7)}_1  = - \fft{\bar\alpha}{\sqrt{1-|\alpha|^2}}(\tilde\psi^{(5)}_2 - \tilde\psi^{(7)}_2).
\end{equation}
In this case we are again left
with the lower component of the dilatino and the upper components of
the gravitino with the exception of $\tilde\psi^{(5)} -
\tilde\psi^{(7)}$ as described above. We proceed again by organizing the spin-1/2 fermions by mass eigenstates,
\begin{eqnarray}
&&\tilde\psi^{m=11/2} = 2(\tilde\psi_1^{(5)} + \tilde\psi_1^{(7)})
+ \tilde\psi_1^{(9)}, \nonumber \\
&&\tilde\psi^{m=-9/2} = \ft12(\tilde\psi_1^{(5)} + \tilde\psi_1^{(7)})
- \tilde\psi_1^{(9)},  \nonumber \\
&&\tilde\psi^{m=-3/2} = \tilde\psi_1^{(5)} - \tilde\psi_1^{(7)}.
\end{eqnarray}
The supersymmetry transformations then become
\begin{eqnarray}\label{eq:BHsusy}
\delta\tilde\lambda&=&\ft{1}{2\tau_2}\gamma\cdot\partial\tau\tilde\varepsilon^c -i\bar\alpha^{-1} e^{-2B}v_i(\gamma\cdot\hat{\bar{\mathcal F}}_1^i - ie^{A-C}\hat{\bar{\mathcal F}}_0^i )\tilde\varepsilon \nonumber \\
\delta\tilde\psi^{m=11/2}&=&\biggl[-\ft{i}2\gamma\cdot\partial(4B+C) - \ft38e^{-4B}\gamma\cdot \mathbb A +\ft18e^{C-A}\gamma\cdot(F_2+e^{-2B-2C}p_2) \nonumber \\
&& - ie^{A-2B+C}-\ft{3i}2(1-|\alpha|^2)^{-1/2}e^{A-C} + \ft{5i}8e^{A-4B-C}(4+\hat\phi) \biggr] \tilde\varepsilon \nonumber \\
&& -\alpha^{-1}e^{-2B}v_i(\ft{3i}4\gamma\cdot\hat{\mathcal{F}}_1^i + \ft74e^{A-C}\hat{\mathcal{F}}_0^i)\tilde\varepsilon^c \nonumber\\
\delta\tilde\psi^{m=-9/2} &=& \Bigl[-\ft{i}2\gamma\cdot(B-C) - \ft14e^{-4B}\gamma\cdot\mathbb A -\ft18e^{C-A}\gamma\cdot(F_2+e^{-2B-2C}p_2) \nonumber \\
&& - \ft{3i}2e^{A-2B+C} + \ft{3i}2(1-|\alpha|^2)^{-1/2}e^{A-C}\Bigr]\tilde\varepsilon + \alpha^{-1}e^{-2B}v_i(-\ft{i}2\gamma\cdot\hat{\mathcal F}_1^i + \ft12e^{A-C}\hat{\mathcal F}_0^i )\tilde\varepsilon^c \nonumber \\
\delta\tilde\psi^{m=-3/2}&=& \ft1{2\sqrt{1-|\alpha|^2}}\Bigl[i\gamma\cdot\left((1-|\alpha|^2)^{-1/2}\Re(\bar\alpha D\alpha) + i \Im(\bar\alpha D\alpha)\right) - 3ie^{A-C}|\alpha|^2\Bigr] \tilde\varepsilon \nonumber \\
&& -\ft12\bar\alpha e^{-2B}v_i\biggl[\gamma\cdot(h_1^i- 2\Im(\alpha f_1^i)) -ie^{A-C}(j_0^i- 2\Im(\alpha f_0^i))\biggr]\tilde\varepsilon^c, \nonumber \\
\delta\hat{\tilde\psi}_\alpha &=&\biggl[D_\alpha+\ft{i}{24}e^{C-A}(\gamma_\alpha{}^{\beta\gamma}
-4\delta_\alpha^\beta\gamma^\gamma)(F_{\beta\gamma}-2e^{-2B-2C}p_{2\,\beta\gamma})\nonumber\\
&&+\ft16\gamma_\alpha\Bigl(2e^{A-2B+C}+3(1-|\alpha|^2)^{-1/2}e^{A-C}-\ft12e^{A-4B-C}(4+\hat\phi_0)\Bigr)
\nonumber\\
&&+\ft{i}{2|\alpha|^2}\left(1-(1-|\alpha|^2)^{-1/2}\right)\Im{(\bar\alpha D_\alpha\alpha)} \biggr]\tilde\varepsilon\nonumber\\
&& +\alpha^{-1}e^{-2B}v_i(\hat{\mathcal F}_\alpha^i  - \ft{i}3\gamma_\alpha e^{A-C}\hat{\mathcal F}_0^i)\tilde\varepsilon^c,
\end{eqnarray}
where we have defined the following combinations
\begin{equation}
\hat{\mathcal F}^i_1 \equiv \Re(\alpha \hat f_1^i) + \ft{i}{\sqrt{1-|\alpha|^2}}\Im(\alpha \hat f_1^i), \qquad \hat{\mathcal F}^i_0 \equiv \Re(\alpha \hat f_0^i) + \ft{i}{\sqrt{1-|\alpha|^2}}\Im(\alpha \hat f_0^i),
\end{equation}
have taken $B=(B_1+B_2)/2$, and have taken the covariant derivative to be
$D_\alpha \equiv \nabla_\alpha - \ft{3i}2(A_\alpha +\ft16e^{-2B_1-2B_2}\mathbb A_\alpha) + \ft{i}{4\tau_2}\partial_\alpha\tau_1 - \ft{3i}2\left(1-(1-|\alpha|^2)^{-1/2}\right)A_\alpha$.
In addition to the universal $\mathcal N=2$ multiplets identified in
\cite{Liu:2010pq}, this reduction retains the Betti hyper-multiplet, with
four real scalars $\{e_0^i,\,\alpha\}$ and a single Dirac fermion corresponding
to the difference $\tilde\psi^{(5)} - \tilde\psi^{(7)}$.  On-shell, these
fields complete an $\mathcal N = 2$ LH+RH chiral multiplet
\begin{eqnarray}
\mathcal D(3,0,0)_2&=&D(3\ft12,\ft12,0)_1+D(3,0,0)_2+D(4,0,0)_0,\nonumber\\
\mathcal D(3,0,0)_{-2}&=&D(3\ft12,\ft12,0)_{-1}+D(3,0,0)_{-2}
+D(4,0,0)_0.
\end{eqnarray}

Comparing this $T^{1,1}$ field content with the spectrum of excitations on $S^5$ \cite{Kim:1985ez}, we see that scalars and a fermion with the appropriate masses show up at the second rung of the KK tower. Curiously, they are the same towers corresponding to the Betti vector modes discussed in the previous section. This seems to indicate that these modes do not belong to a universal consistent truncation, as it would be expected that in a general scenario modes not at the lowest rung of the tower will source modes higher on the tower, thus destroying the consistency.

\subsubsection{$\mathcal N = 2$ superpotential}

Again, in this truncation we can read off an $\mathcal N = 2$ superpotential from the gravitino variation
\begin{equation}\label{eq:BHsuperpot}
W = -\fft12(4+\phi_0)e^{A-4B-C} +2e^{A-2B+C}  + \fft{3}{\sqrt{1-|\alpha|^2}}e^{A-C}.
\end{equation}
This superpotential correctly reproduces the scalar potential according to
(\ref{eq:potspot}), where in this case the inverse scalar metric has a few
more components and is given by
\begin{eqnarray}
(\mathcal G_{\{B,C\}}^{-1})^{ij} = \frac{1}{16}\begin{pmatrix}1&-1\cr-1&7
\end{pmatrix}, \qquad \mathcal G^{-1}_\alpha = -\fft12(1-|\alpha|^2)\begin{pmatrix}\alpha^2&-2+|\alpha|^2\cr-2+|\alpha|^2&\bar\alpha^2
\end{pmatrix}
,&& \\
(\mathcal G_{\{e_0^i,b_0^i,\bar{b}_0^i\}}^{-1})^{ij} =
\fft{e^{4B}}{4(1-|\alpha|^2)}(\mathcal{M}^{-1})^{ij}\begin{pmatrix}2(1+|\alpha|^2)&2i\bar\alpha&-2i\alpha\cr 2i\bar\alpha&\bar\alpha^2&-1\cr-2i\alpha&-1&\alpha^2
\end{pmatrix}, \qquad \mathcal G_{\tau}^{-1} = \tau_2^2.&&
\end{eqnarray}
%

\section{Supersymmetry of the PT ansatz and the KS solution}\label{sec:PTKS}

Most attention on Sasaki-Einstein reductions of IIB and M-theory involving
massive modes has been focused towards applications involving the construction
of duals to condensed matter systems.  However, these constructions are
readily applied to reductions on Calabi-Yau cones.  Consider, for example,
the case of IIB supergravity on a non-compact $CY_3$.  By singling out a
radial direction, and hence writing $CY_3$ as a cone over $SE_5$, we end
up with an effective theory reduced to five dimensions.  The resulting
spacetime metric can be viewed as a warped product of the form
\begin{equation}
ds_5^2 = e^{2X(\rho)}d\rho^2 + e^{2Y(\rho)}h_{\mu\nu}(x)dx^\mu dx^\nu,
\end{equation}
where $h_{\mu\nu}$ is the four-dimensional metric.  (Note that
$X(\rho)$ can be set to zero by an appropriate redefinition of $\rho$.)

For reductions to four-dimensional Minkowski backgrounds, we let
$h_{\mu\nu} = \eta_{\mu\nu}$ and set all non-scalar background fields to
zero.  What remains are scalars, which we take to depend only on the
$\rho$ coordinate.  This is, of course, a standard domain wall ansatz.
Supersymmetric domain wall solutions may be obtained by solving the
resulting Killing spinor equations.  Since the four-dimensional spacetime
is flat, and hence admits trivial parallel spinors, all that remains is
to fix the action of $\gamma^{\bar\rho}$ (where the overline denotes a
tangent space index).  The most general choice is
\begin{eqnarray}\label{eq:ksproj}
\gamma^{\bar\rho} \varepsilon &=& \cos\theta\varepsilon +e^{i\phi}\sin\theta\varepsilon^c, \nonumber \\
\gamma^{\bar\rho} \varepsilon^c &=& e^{-i\phi}\sin\theta\varepsilon-\cos\theta\varepsilon^c,
\end{eqnarray}
where $\theta$ and $\phi$ are two arbitrary (and possibly $\rho$ dependent)
angles.  This is of course a 1/2-BPS projection for the domain wall.

As previously noted in \cite{Bena:2010pr,Cassani:2010na}, the general $\mathcal N = 4$ reduction on $T^{1,1}$ includes the PT ansatz \cite{Papadopoulos:2000gj} as a subset. Unfortunately, the $\mathcal N = 2$ truncations we have focused on only include a further subset of the PT ansatz which contains the deformed conifold solution \cite{Klebanov:2000hb} but not the resolved conifold \cite{Pando Zayas:2000sq} or the Maldacena-Nunez solution \cite{Maldacena:2000yy}. In what follows we will examine the supersymmetry of the Klebanov-Strassler solution from the consistent truncation point of view and make a connection to the previously obtained superpotential. It would be interesting to obtain an $\mathcal N = 4$ superpotential from the full $\mathcal N = 4$ variations given in Section~\ref{sec:N4susy}. However, we leave this for future work.

\subsection{Supersymmetry of the Klebanov-Strassler solution}

As a nontrivial consistency check of our analysis, we examine the Killing
spinor equations of the Klebanov-Strassler solution.  Starting with the
supersymmetry transformations given in Section \ref{sec:bhyper}, this actually
becomes a fairly straightforward exercise. For completeness, we discuss the
precise relation between our ansatz and the KS solution in
Appendix~\ref{sec:KSapp}. The only work we have to do is to obtain the
projection satisfied by the supersymmetry parameter $\tilde\varepsilon$, which,
after examining the solution, turns out to be given by
\begin{equation}\label{eq:ksproj-KS}
\gamma^{\bar\rho} \tilde\varepsilon = -\tilde\varepsilon, \qquad
\gamma^{\bar\rho} \tilde\varepsilon^c = \tilde\varepsilon^c.
\end{equation}
This corresponds to constant $\theta = \pi$ and $\phi = 0$ in (\ref{eq:ksproj}). Using this projection, we can insert the KS solution into the supersymmetry transformations in section \ref{sec:bhyper}. When evaluated on the KS solution, we find that $v_i\hat{\mathcal F}^i_1$ and $v_i\hat{\mathcal F}^i_0$ both vanish, and upon applying the Killing spinor projection the supersymmetry transformations of all the spin-$1/2$ fields vanish.

All that remains is to examine the $\rho$-component of the gravitino variation. We find that the Killing spinor is given simply by
\begin{equation}
\tilde\varepsilon = e^{Y/2} \varepsilon_0,
\end{equation}
where%
\footnote{The precise form of the functions $h(\rho)$ and $K(\rho)$ are given in Appendix \ref{sec:KSapp}.}
\begin{equation}
e^{Y/2} = \left(\fft32\epsilon^{4/3}\right)^{5/12}h^{1/12}K^{1/6}\sinh^{1/3}\rho,
\end{equation}
and $\varepsilon_0$ is now a constant spinor parameter. Note that this is precisely the expected form for the Killing spinor in solutions containing a timelike Killing vector. Using this we can express the ten-dimensional Killing spinor for the Klebanov-Strassler solution, in the notation in (\ref{eq:susyspinor}), as
\begin{equation}
\epsilon = h^{-1/8}e^{\fft{i}{2}\vec\beta\cdot\vec\sigma}
\btop{\varepsilon_0}0,
\end{equation}
in agreement with the literature \cite{Arean:2004mm}.

In addition, we also recover the appropriate superpotential from the general expression given in (\ref{eq:BHsuperpot}).
As a check we map our ansatz to the PT ansatz explicitly%
\footnote{The detailed relation between our ansatz and the ansatz in \cite{Papadopoulos:2000gj} is given in Appendix \ref{sec:PTapp}.}.
Doing so we get the following expression for the superpotential in terms of the fields in \cite{Papadopoulos:2000gj},
\begin{equation}
W = \frac{2^{-2/3}}{3}\Bigl(\fft12 [Q + 2P(h_1+bh_2)] e^{-2x+4p} + e^{-2x-2p} + \cosh y e^{4p}\Bigr),
\end{equation}
which is precisely the superpotential given in \cite{Papadopoulos:2000gj} for the KS ansatz up to a convention dependent overall factor. This again acts as a nontrivial check of our analysis.

\section{Dynamic SU(2) structure in flux compactifications}

Recently an interesting solution of IIB has been constructed in \cite{Heidenreich:2010ad} which possesses dynamic SU(2) structure. This solution is a warped reduction on a cone over $\mathbb{CP}^2$, constructed as the supergravity solution to D7-branes wrapping the internal $\mathbb{CP}^2$ base. A very interesting aspect of this solution is that it contains nontrivial imaginary anti-self-dual (IASD) flux on the cone as a component of the complexified IIB 3-form. In \cite{Baumann:2010sx,Heidenreich:2010ad} it is argued that IASD flux is sourced by gaugino bilinears on the world volume of the D7-branes. Thus gaugino condensation on the world volume will source IASD flux in the supergravity and so this background provides a supergravity description of gaugino condensation.

Another interesting aspect of this solution is that it possesses dynamic SU(2) structure on the cone. The dynamic SU(2) structure is described by a varying inner product between two internal spinors on the six-dimensional manifold. For static SU(2) structure this inner product is constant everywhere. When the inner product varies along the internal manifold the SU(2) structure is termed dynamic. We will find that for our ansatz this dynamic SU(2) structure is encoded in the fact that the five-dimensional killing spinor, $\varepsilon$, satisfies a nontrivial projection which varies along the radial direction of the cone.

The wrapped D7-brane solution of \cite{Heidenreich:2010ad} includes a general presentation of the supersymmetry variations on the cone in the language of generalized complex geometry by utilizing the Killing spinors on the cone. The explicit solution described therein is in a particular limit of this geometry termed the near-stack region, although some properties of the solution for the full cone geometry are also given. The purpose of this section is to relate the full $\mathbb{CP}^2$ cone reduction with the reduction on generic squashed $SE_5$ discussed in \cite{Cassani:2010uw,Liu:2010sa,Gauntlett:2010vu} and whose supersymmetry has been analyzed in \cite{Bah:2010cu,Liu:2010pq}. The theory corresponds to a particular truncation of the $T^{1,1}$ reduction we have analyzed and thus our supersymmetry transformations in Section \ref{sec:N4susy} are still applicable to this reduction after truncating out the additional field content. We present the supersymmetry variations on the full cone as well as discuss the supersymmetry of the near-stack solution found in \cite{Heidenreich:2010ad}.

\subsection{The reduction on the full $\mathbb{CP}^2$ cone}

The full $\mathbb{CP}^2$ ansatz presented in \cite{Heidenreich:2010ad} can be recovered from the general ansatz on $SE_5$. The metric ansatz is given by%
\footnote{To avoid confusion we denote all quantities from \cite{Heidenreich:2010ad} with a subscript $m$. The relation between our ansatz and that of \cite{Heidenreich:2010ad}  can be found in Appendix~\ref{sec:HMTapp}.}
\begin{equation}
ds_{10}^2 = e^{2A_m(\rho)}\eta_{\mu\nu}dx^\mu dx^\nu + e^{-2A_m(\rho)}ds_6(y)^2.
\end{equation}
The internal metric can be brought to the form
\begin{equation}
ds_6^2 = e^{-4B_m}\rho^4[d\rho^2 + \rho^2 (d\psi + \mathcal A_m)^2] + e^{2C_m} ds^2(\mathbb{CP}^2),
\end{equation}
where $ds^2(\mathbb{CP}^2)$ is the Fubini-Study metric on $\mathbb{CP}^2$ and $\mathcal A_m$ is the corresponding Reeb vector. Explicitly these are given by,
\begin{eqnarray}
&&ds^2(\mathbb{CP}^2) = \frac{\delta_{ij}(1+|z_k|^2) - \bar z_i z_j}{(1+|z_k|^2)^2}dz_i d\bar z_j, \nonumber \\
&& \mathcal A_m = \frac{i}{2(1+|z_k|^2)}(z_id\bar z_i - \bar z_i dz_j),
\end{eqnarray}
where $z_i$ are complex coordinates on $\mathbb{CP}^2$. Also, note that the internal manifold is six-dimensional so that the coordinate $\rho$ should be identified as a radial coordinate in the five-dimensional space-time manifold in our reduction as in the discussion at the beginning of Section \ref{sec:PTKS}.

The five-form is described by a single function
\begin{equation}
\tilde F_5 = (1+*_{10})\omega_4\wedge d\alpha_m(\rho),
\end{equation}
where $\omega_4$ is the volume form on the four-dimensional spacetime with metric $ds^2 = \eta_{\mu\nu}dx^\mu dx^\nu$.  In addition, the complex
three-form, which in our notation is $G_{m3} = v_i F_3^i$, may be expanded
in terms of SU(3) invariant forms $\omega_{i,j}$ as
\begin{equation}
G_{m3} = g_{3,0}(\rho)\,\omega_{3,0} + g_{2,1}(\rho)\,\omega_{2,1} + g_{1,2}(\rho)\,\omega_{1,2} + g_{0,3}(\rho)\,\omega_{0,3}.
\end{equation}
In our notation the $\omega_{i,j}$ assume the explicit forms
\begin{eqnarray}
\omega_{3,0} = \rho^2d\rho\wedge\Omega + i \rho^3\Omega\wedge\eta, \qquad \omega_{2,1} = \rho^2d\rho\wedge\Omega - i \rho^3\Omega\wedge\eta, && \nonumber \\
\omega_{0,3} = \rho^2d\rho\wedge\bar\Omega + i \rho^3\bar\Omega\wedge\eta, \qquad \omega_{1,2} = \rho^2d\rho\wedge\bar\Omega - i \rho^3\bar\Omega\wedge\eta.
\end{eqnarray}

The supersymmetry transformations for this reduction can be read off from the general expressions in Section \ref{sec:N4susy}. The vanishing of these conditions should reproduce the expressions given in Section~4.3 of \cite{Heidenreich:2010ad} with the caveat that there they allow for general dependence on the complex fiber.  However we must restrict dependence on this fiber direction to only $\rho$ so that $\partial_\Theta$ reduces as in Eq.~(5.17) in \cite{Heidenreich:2010ad}.

\subsection{Supersymmetry of the near-stack solution}

We now move on to detail the supersymmetry conditions in the near-stack
limit of the geometry described above. While this was already thoroughly
analyzed in \cite{Heidenreich:2010ad}, we nevertheless wish to highlight
the applicability of the squashed Sasaki-Einstein reduction to this system.
The near-stack limit is implemented by scaling the coordinates as
\begin{equation}\label{eq:NSscaling}
z_i \rightarrow \delta\, z_i, \qquad \rho \rightarrow \delta^{1/3} \rho,
\end{equation}
and truncating at a finite order in $\delta$. It will turn out that in order to keep the terms involving IASD flux we will neglect terms of $\mathcal O (\delta^2)$ or higher.

In order to match onto the solution presented in \cite{Heidenreich:2010ad} we must choose the supersymmetry parameter as
\begin{equation}
\varepsilon = \btop{0}{\varepsilon_2}.
\end{equation}
Inserting this into the $\mathcal N=4$ supersymmetry transformations in Section \ref{sec:N4susy}, we arrive at
\begin{eqnarray}
\delta\lambda&=&-\ft{1}{2\tau_2}\gamma\cdot\partial\tau
\varepsilon_2^c + ie^{-2B}v_i\left(\gamma\cdot f_1^i
-ie^{A-C} f_0^i\right)\varepsilon_2, \nonumber \\
\delta\psi^{(5)}&=&\Bigl[-\ft{i}2\gamma\cdot\partial B
+\ft{i}2e^{A-2B+C}+\ft{i}8e^{A-4B-C}(4+\phi_0)\Bigr]\varepsilon_2 \nonumber \\
&&-\ft14v_ie^{-2B}\big(i\gamma\cdot {\bar f}_1^i + e^{A-C} f_0^i\big) \varepsilon_2^c,\nonumber\\
\delta\psi^{(9)}&=&\Bigl[-\ft{i}2\gamma\cdot\partial C
-ie^{A-2B+C} + \ft{3i}2e^{A-C}+\fft
{i}8e^{A-4B-C}(4+\phi_0)\Bigr]\varepsilon_2 \nonumber \\
&&+\ft14v_ie^{-2B}\big(i\gamma\cdot{\bar f}_1^i + 3e^{A-C} {\bar f}_0^i\big) \varepsilon_2^c, \nonumber \\
\delta\psi_\alpha &=& \Bigl[\hat\nabla_\alpha -\ft1{12} \gamma_\alpha\Bigl(4e^{A-2B+C}+6e^{A-C}
+e^{A-4B-C}(4+\hat\phi_0)\Bigr) \Bigr]\varepsilon_2 \nonumber \\
&& + v_ie^{-2B}\big(\bar f^i_\alpha -\ft{i}3e^{A-C}\gamma_\alpha{\bar f^i}_0\big)
\varepsilon_2^c.
\end{eqnarray}
The analysis of these equations in the near-stack region is slightly subtle.
We find it convenient to rewrite the above transformations in terms of the ansatz of \cite{Heidenreich:2010ad}
\begin{eqnarray}
\Lambda(\rho)\delta\lambda&=& -\ft1{2\tau_2}\rho\partial_\rho\tau\gamma^{\bar\rho}\varepsilon_2^c + i\rho^3e^{2A_m}\big((g_{3,0}+g_{2,1})\gamma^{\bar\rho} + (g_{3,0}-g_{2,1})\big)\varepsilon_2 \nonumber \\
\Lambda(\rho)\delta\psi^{(5)}&=&\Bigl(\ft{i}2\gamma^{\bar\rho}\rho\partial_\rho A_m + \ft{i}2\rho^6e^{-4B_m} - \ft{i}8e^{-4A_m}\rho\partial_\rho\alpha_m\Bigr) \varepsilon_2 \nonumber \\
&& - \ft{i}4\rho^3e^{2A_m}\Bigl((g_{0,3}+g_{1,2})\gamma^{\bar\rho} - (g_{0,3}-g_{1,2})\Bigr)\varepsilon_2^c, \nonumber\\
\Lambda(\rho)\delta\psi^{(9)}&=&\Bigl(-\ft{i}2\gamma^{\bar\rho}(3-\rho\partial_\rho A_m - 2\rho\partial_\rho B_m) - i\rho^6e^{-4B_m} + \ft{3i}2 - \ft{i}8e^{-4A_m}\rho\partial_\rho\alpha_m\Bigr) \varepsilon_2 \nonumber \\
&& + \ft{i}4\rho^3e^{2A_m}\Bigl((g_{0,3}+g_{1,2})\gamma^{\bar\rho} + 3(g_{0,3}-g_{1,2})\Bigr)\varepsilon_2^c, \nonumber\\
\rho\delta\psi_\rho &=& \Bigl( \rho\partial_\rho - \ft1{12} \gamma_{\bar\rho}(4\rho^6e^{-4B_m} + 6 - e^{-4A_m}\rho\partial_\rho\alpha_m)\Bigr)\varepsilon_2 \nonumber \\
&&+\rho^3\Bigl((g_{0,3}+g_{1,2}) - \fft13 \gamma_{\bar\rho}(g_{0,3}-g_{1,2})\Bigr)\varepsilon_2^c,
\end{eqnarray}
where $\Lambda(\rho) \equiv \rho^{4}e^{-\ft83A_m - \ft83B_m}$ is an overall $\rho$-dependent factor.

Implementing the near-stack scaling in (\ref{eq:NSscaling}), we see that all terms are $\mathcal O(\delta^0)$, except terms containing $g_{i,j}$, which are $\mathcal O(\delta)$, and the potential term containing $e^{-4B_m}$, which is $\mathcal O(\delta^2)$. In order to keep the IASD flux encoded in $g_{1,2}$ we choose to keep terms of $\mathcal O(\delta)$ or lower, setting $\mathcal O (\delta^2)$ and higher terms to zero.  (We then set $\delta=1$ after performing this scaling.) Finally, choosing the Killing spinor to satisfy,
\begin{equation}
\gamma^{\bar\rho} \varepsilon_2 = \cos\varphi\, \varepsilon_2 + e^{-i\xi}\sin\varphi\, \varepsilon_2^c, \qquad
\gamma^{\bar\rho} \varepsilon_2^c =  e^{i\xi}\sin\varphi\, \varepsilon_2 - \cos\varphi\, \varepsilon_2^c,
\end{equation}
and using the supersymmetry conditions derived in \cite{Heidenreich:2010ad}, which are given in (\ref{eq:HMTsusycond}), it can be seen that the above supersymmetry transformations all vanish provided
\begin{equation}
\varepsilon_2 = e^{Y/2}\varepsilon_0,
\end{equation}
with $\varepsilon_0$ a constant spinor. In our notation the dynamic SU(2) structure is encoded in the above expression and in the Killing spinor projection, where the parameter $\varphi$ is a nontrivial function of $\rho$.

One detail that we have glossed over in the above discussion involves the limiting procedure to the near-stack geometry. In particular, we find that dropping the potential term proportional to $\rho^6e^{-4B_m}$ was needed in order to satisfy the supersymmetry conditions. The solution for $B_m$ arrived at in this way yields
\begin{equation}
\rho^6e^{-4B_m} = \frac{9|c_1|^2}{\sin^2\varphi}e^{-4A_m},
\end{equation}
which is of the same order as the other terms in the variation, thus implying that the exclusion of this term is seemingly not justified. In fact, this precise term was discussed in Section 6.3 of \cite{Heidenreich:2010ad} as arising from effects due to the D7-branes acting as sources in the full ten-dimensional geometry as opposed to isolating the near-stack region, where they appear to only source modes within the cone. The resolution to this given in \cite{Heidenreich:2010ad} is to impose a further separation of scales such that
\begin{equation}
\rho^6e^{-4B_m} \ll 1.
\end{equation}

Evaluating the supersymmetry conditions in the near-stack region supplemented with the above condition then gives precise agreement with \cite{Heidenreich:2010ad}. This highlights the point made in \cite{Heidenreich:2010ad} that the solution obtained is appropriate for a finite range of the coordinate $\rho$ and corrections to this solution are parametrically suppressed within this region.%
\footnote{We thank L. McAllister and B. Heidenreich for clarifications on this.}
This discussion seems to highlight the fact that, as is done in \cite{Heidenreich:2010ad}, care must be taken when interpreting approximate solutions and emphasizes the desire to obtain exact solutions when possible. The approximate solution obtained in this way nevertheless is an interesting first step towards understanding solutions in these backgrounds.

\section{Discussion}

We have presented an analysis of the fermionic supersymmetry transformations for
reductions of IIB supergravity on squashed $T^{1,1}$ containing massive modes.

We presented the full $\mathcal N = 4$ theory, including a detailed exposition of the supersymmetry transformations and equations of motion. Following this we focused on the supersymmetry variations of truncations to $\mathcal N = 2$ supersymmetry, where we were able to read off the relevant superpotentials. This analysis furthermore allows us
to investigate the Killing spinors of cones over $T^{1,1}$.  As an example, we
examined the Klebanov-Strassler solution and also recovered the superpotential given in \cite{Papadopoulos:2000gj}. The knowledge of the generic superpotentials for the $\mathcal N = 2$ reductions may prove useful in determining other supersymmetric solutions.

While we have only constructed superpotentials for $\mathcal N = 2$ truncations, a more thorough analysis of the entire $\mathcal N = 4$ theory is required in order to make contact with the entire PT ansatz. In particular, it would be interesting to determine the superpotential describing the full $\mathcal N = 4$ theory. This would be desirable as it would also allow a direct connection with interpolating solutions such as those described in \cite{Butti:2004pk}, and may help to further classify four and five-dimensional solutions of IIB supergravity. We leave the study of such a superpotential to future work.

Given the success of extending the generic consistent truncation on $SE_5$ to the case of
$T^{1,1}$, one may wonder whether a similar Betti vector reduction can be obtained for
the more general class of $Y^{p,q}$ manifolds which also admit U(1) structure.  Unfortunately,
this possibility seems unlikely, as while a closed anti-self-dual form $J_-$ does exist, it
however does not square to the volume form on the base.  In particular
\begin{equation}
J_-\wedge J_-=-\fft2{(1-y)^4}*_41,
\end{equation}
in the notation of \cite{Gauntlett:2004yd}.  The explicit dependence on the internal
coordinate $y$ spoils the
consistency of the truncation retaining the Betti vector modes associated with $J_-$.

Finally, we also demonstrated how the consistent massive truncations of IIB supergravity on $SE_5$ manifolds may be related to a class of flux compactifications on Calabi-Yau cones. Much of the recent work on flux compactifications has utilized the powerful machinery of generalized geometry. We hope that, in relating some of these types of compactifications to the consistent truncations we have been discussing, it might be possible to elucidate some of the structure of these types of supergravity solutions as well as to provide a common ground between the two approaches.

\begin{acknowledgments}

We wish to thank I.~Bah, I.~Bena, A.~Faraggi, M.~Gra\~na, N.~Halmagyi, B.~Heidenreich,
L.~McAllister and L.~Pando Zayas for useful discussions.  Initial development of the $T^{1,1}$
reduction ansatz was done in collaboration with N.~Halmagyi. JTL wishes to
acknowledge the hospitality of the LPTHE Jussieu and IPhT CEA/Saclay where
this project was initiated. This work was supported in part by the US
Department of Energy under grant DE-FG02-95ER40899.

\end{acknowledgments}

\appendix

\section{Relation between hatted and un-hatted quantities}\label{sec:shiftedapp}

For convenience, we define both the unshifted tensors
\begin{equation}
J_1=\fft{i}{12}E_1\wedge\bar{E}_1,\qquad
J_2=\fft{i}{12}E_2\wedge\bar{E}_2,\qquad
\Omega=\fft16E_1\wedge E_2,
\end{equation}
and the shifted tensors
\begin{equation}
\hat J_1=\fft{i}{12}\hat E_1\wedge\hat{\bar{E}}_1,\qquad
\hat J_2=\fft{i}{12}\hat E_2\wedge\hat{\bar{E}}_2,\qquad
\hat\Omega=\fft16\hat E_1\wedge\hat E_2.
\end{equation}
The relation between these sets of tensors is given by
\begin{equation}
\hat J_1=J_1,\qquad
\hat J_2=J_2-|\alpha|^2J_1+\Im(\bar\alpha\Omega),\qquad
\hat\Omega=\Omega-2i\alpha J_1,
\end{equation}
and the inverse
\begin{equation}
J_1=\hat J_1,\qquad
J_2=\hat J_2-|\alpha|^2\hat J_1-\Im(\bar\alpha\hat\Omega),\qquad
\Omega=\hat\Omega+2i\alpha\hat J_1.
\end{equation}
Expanding the three forms in terms of the shifted tensors, we have
\begin{eqnarray}
F_3^i&=&\hat g_3^i+\hat g_2^i\wedge(\eta+A_1)+\hat g_{11}^i\wedge\hat J_1
+\hat g_{12}^i\wedge\hat J_2+\hat j_{01}^i\hat J_1\wedge(\eta+A_1)
+\hat j_{02}\hat J_2\wedge(\eta+A_1)\nonumber\\
&&+2\Re[\hat f_1^i\wedge\hat\Omega
+\hat f_0^i\hat\Omega\wedge(\eta+A_1)],
\end{eqnarray}
with the following relations
\begin{eqnarray}
\hat g_3^i&=&g_3^i,\nonumber\\
\hat g_2^i&=&g_2^i,\nonumber\\
\hat g_{11}^i&=&(1-|\alpha|^2)g_1^i+(1+|\alpha|^2)h_1^i+2i\alpha f_1^i
-2i\overline\alpha\bar f_1^i,\nonumber\\
\hat g_{12}^i&=&g_1^i-h_1^i,\nonumber\\
\hat f_1^i&=&f_1^i+\ft12i\bar\alpha(g_1^i-h_1^i),\nonumber\\
\hat f_0^i&=&f_0^i-\ft12i\bar\alpha j_0^i,\nonumber\\
\hat j_{01}^i&=&(1+|\alpha|^2)j_0^i+2i\alpha f_0^i
-2i\bar\alpha\bar f_0^i,\nonumber\\
\hat j_{02}^i&=&-j_0^i.
\end{eqnarray}
Likewise, for the five-form, we have
\begin{eqnarray}
\widetilde F_5&=&(4+\hat\phi_0)*_41\wedge(\eta+A_1)+\hat{\mathbb A}_1\wedge*_41
+\hat p_{21}\wedge\hat J_1\wedge(\eta+A_1)+\hat p_{22}\wedge\hat J_2\wedge
(\eta+A_1)\nonumber\\
&&+2\Re(\hat q_2\wedge\hat\Omega\wedge(\eta+A_1))
+e^{5A-2B_1-2B_2-C}(4+\hat\phi_0)*1\nonumber\\
&&-e^{3A-2B_1-2B_2+C}*\hat{\mathbb A}_1\wedge(\eta+A_1)
+e^{A-2B_1+2B_2-C}*\hat p_{21}\wedge\hat J_2\nonumber\\
&&+e^{A+2B_1-2B_2-C}*\hat p_{22}\wedge\hat J_1
+2\Re(e^{A-C}*\hat q_2\wedge\hat\Omega),
\end{eqnarray}
where
\begin{eqnarray}
\hat\phi_0&=&\phi_0,\nonumber\\
\hat{\mathbb A}_1&=&\mathbb A_1,\nonumber\\
\hat p_{21}&=&(1-|\alpha|^2)p_2+(1+|\alpha|^2)r_2+2i\alpha q_2
-2i\bar\alpha\bar q_2,\nonumber\\
\hat p_{22}&=&p_2-r_2,\nonumber\\
\hat q_2&=&q_2+\ft12i\bar\alpha(p_2-r_2).
\end{eqnarray}
%


\section{Reduction of the covariant derivative}\label{app:covderiv}

The ten-dimensional covariant derivative acting on spin-1/2 and spin-3/2 fermions is given by
\begin{eqnarray}
\nabla_M\lambda&=&[\partial_M+\ft14\omega_M{}^{AB}\Gamma_{AB}]\lambda,\nn\\
\nabla_M\psi^A&=&[\partial_M+\ft14\omega_M{}^{BC}\Gamma_{BC}]\psi^A+\omega_M{}^A{}_B
\psi^B.
\label{eq:nabladef}
\end{eqnarray}
Defining $\nabla_M^{(0)}\equiv\partial_M+\fft14\omega_M{}^{AB}\Gamma_{AB}$, we find
\begin{eqnarray}
\nabla^{{(0)}}_\alpha &=& e^{-A}\Big[\nabla^5_\alpha + \ft12 \gamma_\alpha{}^{\beta}\partial_\beta
A - \ft34 A_\alpha(\tilde\gamma^{56}+\tilde\gamma^{78})
+\ft{i}4\hat\sigma_3e^{-A+C}\gamma^\beta F_{\alpha\beta}\tilde\gamma^9\nn\\
&&\kern3em+\ft14 e^{-B_1+B_2}\tilde\gamma^{57}D_\alpha(\alpha_1+\alpha_2\tilde\gamma^{56})
(1-\tilde\gamma^{56}\tilde\gamma^{78})\Big],
\end{eqnarray}
along with
\begin{eqnarray}
\tilde\gamma^5\nabla^{(0)}_5&=&e^{-A}\Big[-\ft{i}2\hat\sigma_3\gamma^\alpha\partial_\alpha B_1
+\ft12e^{A-2B_1+C}(1-|\alpha|^2)\tilde\gamma^{56}\tilde\gamma^9
-\ft{i}4\hat\sigma_3e^{-B_1+B_2}\tilde\gamma^{57}\gamma^\alpha
D_\alpha(\alpha_1+\alpha_2\tilde\gamma^{78})  \nn\\
&& \kern3em-\tilde\gamma^{57}(\ft12e^{A-B_1-B_2+C} + \ft34e^{A-B_1+B_2-C})
(\alpha_1+\alpha_2\tilde\gamma^{78})\tilde\gamma^{78}\tilde\gamma^9\Big]\nn \\
&&+\ft{\sqrt6}2e^{-B_1}\cot\theta_1\sin(\psi/2)(\tilde\gamma^{56}-\tilde\gamma^{78}),\nn\\
\tilde\gamma^6\nabla^{(0)}_6 &=& e^{-A}\Big[-\ft{i}2\hat\sigma_3\gamma^\alpha\partial_\alpha B_1
+\ft12e^{A-2B_1+C}(1-|\alpha|^2)\tilde\gamma^{56}\tilde\gamma^9
+\ft{i}4\hat\sigma_3e^{-B_1+B_2}\tilde\gamma^{68}\gamma^\alpha
D_\alpha(\alpha_1+\alpha_2\tilde\gamma^{78})\nn\\
&&\kern3em-\tilde\gamma^{68}(\ft12e^{A-B_1-B_2+C} + \ft34e^{A-B_1+B_2-C})
(\alpha_1+\alpha_2\tilde\gamma^{78})\tilde\gamma^{78}\tilde\gamma^9\Big]\nn \\
&&-\ft{\sqrt6}2e^{-B_1}\cot\theta_1\cos(\psi/2)(\tilde\gamma^{56}-\tilde\gamma^{78}),\nn\\
\tilde\gamma^7\nabla^{(0)}_7&=&e^{-A}\Big[-\ft{i}2\hat\sigma_3\gamma^\alpha\partial_\alpha B_2
+\ft12e^{A-2B_2+C}\tilde\gamma^{78}\tilde\gamma^9
+\ft{i}4\hat\sigma_3e^{-B_1+B_2}\tilde\gamma^{57}\gamma^\alpha
D_\alpha(\alpha_1+\alpha_2\tilde\gamma^{56})\nn\\
&&\kern3em-\tilde\gamma^{57}(\ft12e^{A-B_1-B_2+C} - \ft34e^{A-B_1+B_2-C})
(\alpha_1+\alpha_2\gamma^{56})\tilde\gamma^{56}\tilde\gamma^9\Big]\nn \\
&&-\ft{\sqrt6}2e^{-B_2}\cot\theta_2\sin(\psi/2)(\tilde\gamma^{56}-\tilde\gamma^{78}),\nn\\
\tilde\gamma^8\nabla^{(0)}_8&=&e^{-A}\Big[-\ft{i}2\hat\sigma_3\gamma^\alpha\partial_\alpha B_2
+\ft12e^{A-2B_2+C}\tilde\gamma^{78}\tilde\gamma^9
-\ft{i}4\hat\sigma_3 e^{-B_1+B_2}\tilde\gamma^{68}\gamma^\alpha
D_\alpha(\alpha_1 + \alpha_2\gamma^{56}) \nn\\
&&\kern3em-\tilde\gamma^{68}(\ft12e^{A-B_1-B_2+C} - \ft34e^{A-B_1+B_2-C})
(\alpha_1+ \alpha_2 \tilde\gamma^{56})\tilde\gamma^{56}\tilde\gamma^9\Big]\nn \\
&&+\ft{\sqrt6}2e^{-B_2}\cot\theta_2\cos(\psi/2)(\tilde\gamma^{56}-\tilde\gamma^{78}),\nn\\
\tilde\gamma^9 \nabla^{(0)}_9 &=& e^{-A}\Big[-\ft{i}2\hat\sigma_3\gamma^\alpha\partial_\alpha C
-\ft18e^{-A+C}\gamma^{\alpha\beta}F_{\alpha\beta}\tilde\gamma^9
+\ft34 e^{A-C}(\tilde\gamma^{56}+\tilde\gamma^{78})\tilde\gamma^9 \nn \\
&&\kern3em-\ft12 e^{A+C}(e^{-2B_1}(1-|\alpha|^2)\tilde\gamma^{56}+e^{-2B_2}\tilde\gamma^{78})
\tilde\gamma^9\nn \\
&&\kern3em+\tilde\gamma^{57}(\ft12e^{A-B_1-B_2+C}-\ft34e^{A-B_1+B_2-C})
(\alpha_1+\alpha_2\tilde\gamma^{56})(\tilde\gamma^{56}+\tilde\gamma^{78})\tilde\gamma^9\Big].
\end{eqnarray}
Note that the factor $(\tilde\gamma^{56}-\tilde\gamma^{78})
=\tilde\gamma^{56}(1+\tilde\gamma^9)$ vanishes when acting on a Killing spinor $\eta$, so the
angle dependent factors do not show up in covariant derivatives of $\lambda$ and $\epsilon$.
Furthermore, they will be cancelled by the additional spin connection term in (\ref{eq:nabladef})
when computing the covariant derivative of the gravitino.

By demanding the action of the above covariant derivatives on the Killing spinor vanish in the simple ``un-deformed" $T^{1,1}$ reduction we can deduce the projections satisfied by the Killing spinor. One finds that when acting directly on a Killing spinor pair $[\begin{matrix}\eta&\eta^c\end{matrix}]^T$, the
internal Dirac matrices take on the eigenvalues
\begin{equation}
\tilde\gamma^{56}=\tilde\gamma^{78}=i\sigma_3,\qquad\tilde\gamma^9=-1.
\end{equation}
For $\tilde\gamma^{57}$, we have
\begin{equation}
\tilde\gamma^{57}=\ft18[\tilde\gamma\cdot\hat\Omega+\tilde\gamma\cdot\hat{\bar\Omega}],\qquad
\tilde\gamma^{57}\sigma_3=\ft18[-\tilde\gamma\cdot\hat\Omega+\tilde\gamma\cdot\hat{\bar\Omega}].
\end{equation}
Using
\begin{equation}
\ft18\tilde\gamma\cdot\hat\Omega=-\sigma_+,\qquad
\ft18\tilde\gamma\cdot\hat{\bar\Omega}=\sigma_-,
\end{equation}
this is equivalent to
\begin{equation}
\tilde\gamma^{57}=-i\sigma_2,\qquad\tilde\gamma^{57}\sigma_3=\sigma_1.
\end{equation}
%


\section{Relations to other solutions}

\subsection{Klebanov-Strassler}
\label{sec:KSapp}

The Klebanov-Strassler ansatz \cite{Klebanov:2000hb} in a notation which makes comparison to ours fairly straightforward is,
\begin{eqnarray}
ds_{10}^2 &=& h^{-1/2}\biggl[\fft{\epsilon^{4/3}h}{6K^2} d \rho^2 + dx_\mu dx^\mu\biggr] \nonumber \\
&& + \fft12fh^{1/2}\epsilon^{4/3}K\biggl[\fft12\cosh 2 \rho\sum^{4}_{i=1}(e^i)^2 + e^1e^3+e^2e^4 + \fft1{3K^3}(e^5)^2\biggr],\nonumber \\
C_2 &=& 3M(F-\fft12)(\Omega +\bar\Omega),\nonumber \\
F_3 &=& dC_2 - 3M e^5\wedge J_-,\nonumber \\
B_2 &=& -3g_s M (f+k)J_- + \fft{3i}2 g_s M (f-k)(\Omega-\bar\Omega),\nonumber \\
H_3 &=& dB_2 \nonumber,\\
\tilde F_5 &=& B_2\wedge F_3 = -108g_s M^2 \ell*_41\wedge\eta,
\end{eqnarray}
where
\begin{equation}
K(\rho) = \frac{(\sinh(2 \rho)-2 \rho)^{1/3}}{2^{1/3}\sinh \rho}.
\end{equation}
To relate this to our ansatz and to satisfy the supersymmetry equations, it turns out to be necessary to perform a world-sheet parity transformation on the IIB fields above,
\begin{equation}
\{B_2, \tilde F_5 \} \rightarrow \{-B_2, -\tilde F_5\},
\end{equation}
with all other fields in the solution invariant. After performing this parity transformation the Klebanov-Strassler solution is recovered from our ansatz by the following identifications, for the three and five-forms,
\begin{eqnarray}
&& 4+ \phi_0 = 108g_sM^2\ell, \qquad e_0^1 =  3g_sM(f+k), \qquad j_0^2 = - 9 M, \nonumber \\
&& b_0^1 = -\fft{3i}2g_sM(f-k), \qquad b_0^2 = 3M(F-\fft12),
\end{eqnarray}
and for the terms in the metric,
\begin{eqnarray}
\alpha_1 &=& \cosh^{-1}\rho, \nonumber \\
e^{2B} &=& \fft32h^{1/2}\epsilon^{4/3}K\sinh\rho, \nonumber \\
e^{2C} &=& \fft32h^{1/2}\epsilon^{4/3}K^{-2}, \nonumber \\
e^{2X} &=& \fft14\biggl(\fft32\biggr)^{2/3}\epsilon^{32/9}h^{4/3}K^{-4/3}\sinh^{4/3}\rho,\nonumber \\
e^{2Y} &=& \biggl(\fft32\biggr)^{5/3}\epsilon^{20/9}h^{1/3}K^{2/3}\sinh^{4/3}\rho,
\end{eqnarray}
where, $\ell = f(1-F) +kF$, with all other fields in our ansatz set to zero. The solution to the above ansatz given in \cite{Klebanov:2000hb} is then,
\begin{eqnarray}
f&=& \frac{ \rho\coth  \rho -1}{2\sinh  \rho}(\cosh \rho-1), \nonumber \\
k&=& \frac{ \rho\coth  \rho -1}{2\sinh  \rho}(\cosh \rho+1), \nonumber \\
F&=& \frac{\sinh \rho- \rho}{2\sinh \rho},
\end{eqnarray}
and the warp factor is given in differential form,
\begin{equation}
h' = - 4\biggl(\fft23\biggr)^{2/3}\frac{g_sM^2\ell}{K^2\sinh^2 \rho}.
\end{equation}

\subsection{PT ansatz}\label{sec:PTapp}

The generalized ansatz of Papadopoulos and Tseytlin \cite{Papadopoulos:2000gj} is also contained within our framework. The following mapping translates between their notation and ours,
\begin{eqnarray}
&& e^{2B_1} = 6e^{x+g}, \qquad e^{2B_2} = 6e^{x-g}, \qquad e^{2C} = 9e^{-6p-x}, \nonumber \\
&&\alpha_1 = -a, \qquad \alpha_2 = 0, \qquad c_0^1 = 6\chi, \qquad c_0^2 = 0, \qquad e_0^1 = 6h_1, \nonumber \\
&&e_0^2 = 0, \qquad j_0^1 = 0, \qquad j_0^2 = 18P, \qquad b_0^1 = 3ih_2, \qquad b_0^2 = -3 P b, \nonumber \\
&&(4+\phi_0) = -108K = -108[Q + 2P(h_1+bh_2)].
\end{eqnarray}

\subsection{Warped D7-brane solution of \cite{Heidenreich:2010ad}}\label{sec:HMTapp}

The map between the full ansatz on the $\mathbb{CP}^2$ cone in \cite{Heidenreich:2010ad} and ours is given by:
\begin{eqnarray}
&&e^{2B} = e^{-2A_m+2C_m}, \qquad e^{2C} = \rho^6 e^{-2A_m-4B_m}, \nonumber \\
&& e^{2X} = \rho^6e^{-8/3(2A_m+2B_m-C_m)}, \qquad e^{2Y} = \rho^2e^{-4/3(A_m+B_m-2C_m)},
\end{eqnarray}
for the metric factors. For the five form,
\begin{equation}
e^{4Y+X+5A-4B-C}(4+\phi_0) = - \alpha_m'.
\end{equation}
And for the three form,
\begin{eqnarray}
&& v_if_1^i = (g_{3,0}+g_{2,1})\rho^2d\rho, \qquad v_if_0^i = i\rho^3(g_{3,0}-g_{2,1}), \nonumber \\
&& v_i\bar{f}_1^i = (g_{0,3}+g_{1,2})\rho^2d\rho, \qquad v_i\bar{f}_0^i = -i\rho^3(g_{0,3}-g_{1,2}).
\end{eqnarray}

The supersymmetry conditions in the near stack region result in the following relations,
\begin{eqnarray}\label{eq:HMTsusycond}
&& g_{3,0} + g_{2,1} = \fft{i\partial_r\tau}{2\tau_2}\cot\varphi\, e^{-2A_m+i\xi}, \qquad g_{3,0} - g_{2,1}= -\fft{i\partial_r\tau}{2\tau_2} e^{-2A_m+i\xi}, \nonumber \\
&& g_{0,3} + g_{1,2} = \fft12\partial_r\varphi e^{-2A_m-i\xi}, \qquad g_{0,3} - g_{1,2} = - \fft12(4\sin\varphi \partial_r A_m + \cos\varphi \partial_r\varphi)e^{-2A_m-i\xi}, \nonumber \\
&& \alpha = \cos\varphi\,e^{4A_m} , \qquad B_m = A_m -\fft14 \log\Bigl[\fft{|c_1|^2}{r^2\sin^2\varphi}\Bigr],
\end{eqnarray}
where $r=\ft13\rho^3$.  Note also that $C_m$ is taken to be a constant.



\begin{thebibliography}{99}


\bibitem{Bremer:1998zp}
M.~S.~Bremer, M.~J.~Duff, H.~Lu, C.~N.~Pope and K.~S.~Stelle,
{\sl Instanton cosmology and domain walls from M-theory and string theory},
Nucl.\ Phys.\  B {\bf 543}, 321 (1999) [arXiv:hep-th/9807051].

\bibitem{Liu:2000gk}
J.~T.~Liu and H.~Sati,
{\sl Breathing mode compactifications and supersymmetry of the brane-world},
Nucl.\ Phys.\  B {\bf 605}, 116 (2001) [arXiv:hep-th/0009184].

\bibitem{Gauntlett:2009zw}
J.~P.~Gauntlett, S.~Kim, O.~Varela and D.~Waldram,
{\sl Consistent supersymmetric Kaluza--Klein truncations with massive modes},
JHEP {\bf 0904}, 102 (2009) [arXiv:0901.0676 [hep-th]].

\bibitem{Cassani:2010uw}
D.~Cassani, G.~Dall'Agata and A.~F.~Faedo,
{\sl Type IIB supergravity on squashed Sasaki-Einstein manifolds},
JHEP {\bf 1005}, 094 (2010) [arXiv:1003.4283 [hep-th]].

\bibitem{Liu:2010sa}
J.~T.~Liu, P.~Szepietowski, Z.~Zhao,
{\sl Consistent massive truncations of IIB supergravity on Sasaki-Einstein manifolds},
Phys.\ Rev.\  {\bf D81}, 124028 (2010) [arXiv:1003.5374 [hep-th]].

\bibitem{Gauntlett:2010vu}
J.~P.~Gauntlett and O.~Varela,
{\sl Universal Kaluza-Klein reductions of type IIB to $\mathcal N=4$
supergravity in five dimensions},
JHEP {\bf 1006}, 081 (2010) [arXiv:1003.5642 [hep-th]].

\bibitem{Skenderis:2010vz}
K.~Skenderis, M.~Taylor and D.~Tsimpis,
{\sl A consistent truncation of IIB supergravity on manifolds admitting a
Sasaki-Einstein structure},
JHEP {\bf 1006}, 025 (2010) [arXiv:1003.5657 [hep-th]].

\bibitem{Cassani:2010na}
D.~Cassani and A.~F.~Faedo,
{\sl A supersymmetric consistent truncation for conifold solutions},
Nucl.\ Phys.\ B {\bf 843}, 455 (2011) [arXiv:1008.0883 [hep-th]].

\bibitem{Bena:2010pr}
I.~Bena, G.~Giecold, M.~Grana, N.~Halmagyi and F.~Orsi,
{\sl Supersymmetric Consistent Truncations of IIB on $T^{1,1}$},
JHEP {\bf 1104}, 021 (2011) [arXiv:1008.0983 [hep-th]].

\bibitem{Cassani:2009ck}
D.~Cassani, A.~-K.~Kashani-Poor,
{\sl Exploiting N=2 in consistent coset reductions of type IIA},
Nucl.\ Phys.\  {\bf B817}, 25-57 (2009) [arXiv:0901.4251 [hep-th]].

\bibitem{Papadopoulos:2000gj}
G.~Papadopoulos, A.~A.~Tseytlin,
{\sl Complex geometry of conifolds and five-brane wrapped on two sphere},
Class.\ Quant.\ Grav.\  {\bf 18}, 1333-1354 (2001) [hep-th/0012034].

\bibitem{Klebanov:2000hb}
I.~R.~Klebanov, M.~J.~Strassler,
{\sl Supergravity and a confining gauge theory: Duality cascades and $\chi$SB resolution of naked singularities},
JHEP {\bf 0008}, 052 (2000) [hep-th/0007191].

\bibitem{Maldacena:2000yy}
J.~M.~Maldacena, C.~Nunez,
{\sl Towards the large N limit of pure N=1 superYang-Mills},
Phys.\ Rev.\ Lett.\  {\bf 86}, 588-591 (2001) [hep-th/0008001].

\bibitem{Pando Zayas:2000sq}
L.~A.~Pando Zayas, A.~A.~Tseytlin,
{\sl 3-branes on resolved conifold},
JHEP {\bf 0011}, 028 (2000) [hep-th/0010088].

\bibitem{Heidenreich:2010ad}
B.~Heidenreich, L.~McAllister, G.~Torroba,
{\sl Dynamic SU(2) Structure from Seven-branes},
JHEP {\bf 1105}, 110 (2011) [arXiv:1011.3510 [hep-th]].

\bibitem{Halmagyi:2011yd}
N.~Halmagyi, J.~T.~Liu and P.~Szepietowski,
{\sl On $\mathcal N = 2$ Truncations of IIB on $T^{1,1}$},
arXiv:1111.6567 [hep-th].

\bibitem{Bah:2010yt}
I.~Bah, A.~Faraggi, J.~I.~Jottar, R.~G.~Leigh and L.~A.~P.~Zayas,
{\sl Fermions and $D=11$ Supergravity On Squashed Sasaki-Einstein Manifolds},
JHEP {\bf 1102}, 068 (2011) [arXiv:1008.1423 [hep-th]].

\bibitem{Bah:2010cu}
I.~Bah, A.~Faraggi, J.~I.~Jottar and R.~G.~Leigh,
{\sl Fermions and Type IIB Supergravity On Squashed Sasaki-Einstein Manifolds},
JHEP {\bf 1101}, 100 (2011) [arXiv:1009.1615 [hep-th]].

\bibitem{Liu:2010pq}
J.~T.~Liu, P.~Szepietowski, Z.~Zhao,
{\sl Supersymmetric massive truncations of IIB supergravity on Sasaki-Einstein manifolds},
Phys.\ Rev.\  {\bf D82}, 124022 (2010) [arXiv:1009.4210 [hep-th]].

\bibitem{Baumann:2010sx}
D.~Baumann, A.~Dymarsky, S.~Kachru, I.~R.~Klebanov and L.~McAllister,
{\sl D3-brane Potentials from Fluxes in AdS/CFT},
JHEP {\bf 1006}, 072 (2010) [arXiv:1001.5028 [hep-th]].

\bibitem{Koerber:2007xk}
P.~Koerber, L.~Martucci,
{\sl From ten to four and back again: How to generalize the geometry},
JHEP {\bf 0708}, 059 (2007) [arXiv:0707.1038 [hep-th]].

\bibitem{Dymarsky:2010mf}
A.~Dymarsky, L.~Martucci,
{\sl D-brane non-perturbative effects and geometric deformations},
JHEP {\bf 1104}, 061 (2011) [arXiv:1012.4018 [hep-th]].

\bibitem{Schwarz:1983qr}
J.~H.~Schwarz,
{\sl Covariant Field Equations Of Chiral N=2 D=10 Supergravity},
Nucl.\ Phys.\  B {\bf 226}, 269 (1983).

\bibitem{Kim:1985ez}
H.~J.~Kim, L.~J.~Romans and P.~van Nieuwenhuizen,
{\sl The Mass Spectrum Of Chiral $N=2$ $D=10$ Supergravity On $S^5$},
Phys.\ Rev.\  D {\bf 32}, 389 (1985).

\bibitem{Ceresole:1999zs}
A.~Ceresole, G.~Dall'Agata, R.~D'Auria and S.~Ferrara,
{\sl Spectrum of type IIB supergravity on AdS$_5\times T^{11}$:
Predictions on $\mathcal N = 1$ SCFT's},
Phys.\ Rev.\  D {\bf 61}, 066001 (2000) [hep-th/9905226].

\bibitem{Ceresole:1999ht}
A.~Ceresole, G.~Dall'Agata and R.~D'Auria,
{\sl KK spectroscopy of type IIB supergravity on AdS$_5\times T^{11}$},
JHEP {\bf 9911}, 009 (1999) [hep-th/9907216].

\bibitem{Arean:2006nc}
D.~Arean,
{\sl Killing spinors of some supergravity solutions},
hep-th/0605286.

\bibitem{Arean:2004mm}
D.~Arean, D.~E.~Crooks and A.~V.~Ramallo,
{\sl Supersymmetric probes on the conifold},
JHEP {\bf 0411}, 035 (2004) [hep-th/0408210].

\bibitem{Butti:2004pk}
A.~Butti, M.~Grana, R.~Minasian {\it et al.},
{\sl The Baryonic branch of Klebanov-Strassler solution: A supersymmetric family of SU(3) structure backgrounds},
JHEP {\bf 0503}, 069 (2005) [hep-th/0412187].

\bibitem{Gauntlett:2004yd}
J.~P.~Gauntlett, D.~Martelli, J.~Sparks and D.~Waldram,
{\sl Sasaki-Einstein metrics on $S^2\times S^3$},
Adv.\ Theor.\ Math.\ Phys.\  {\bf 8}, 711 (2004) [arXiv:hep-th/0403002].

\end{thebibliography}
\end{document}